\documentclass[aps,prd,showpacs,eqsecnum,twocolumn,superscriptaddress,nofootinbib]
{revtex4-1}
\usepackage{amsmath,amssymb,graphicx,color}

\usepackage{color}

\begin{document}

\title{Measurability of the tidal deformability by gravitational waves
  from coalescing binary neutron stars}

\author{Kenta Hotokezaka} \affiliation{Racah Institute of Physics, The
Hebrew University of Jerusalem, Jerusalem, 91904, Israel}

\author{Koutarou Kyutoku} \affiliation{Interdisciplinary Theoretical
Science (iTHES) Research Group, RIKEN, Wako, Saitama 351-0198, Japan}


\author{Yu-ichiro Sekiguchi} \affiliation{Department of Physics, Toho
University, Funabashi, Chiba 274-8510, Japan}

\author{Masaru Shibata}
\affiliation{Yukawa Institute for Theoretical Physics, 
Kyoto University, Kyoto, 606-8502, Japan} 

\date{\today}

\newcommand{\beq}{\begin{equation}}
\newcommand{\eeq}{\end{equation}}
\newcommand{\beqn}{\begin{eqnarray}}
\newcommand{\eeqn}{\end{eqnarray}}
\newcommand{\pa}{\partial}
\newcommand{\vp}{\varphi}
\newcommand{\varep}{\varepsilon}
\newcommand{\ep}{\epsilon}
\newcommand{\comp}{(M/R)_\infty}
\begin{abstract}

Combining new gravitational waveforms derived by long-term (14--16
orbits) numerical-relativity simulations with waveforms by an
effective-one-body (EOB) formalism for coalescing binary neutron
stars, we construct hybrid waveforms and estimate the measurability
for the dimensionless tidal deformability of the neutron stars,
$\Lambda$, by advanced gravitational-wave detectors. We focus on the
equal-mass case with the total mass $2.7M_\odot$. We find that for an
event at a hypothetical effective distance of $D_{\rm eff}=200$\,Mpc,
the distinguishable difference in the dimensionless tidal
deformability will be $\approx 100$, 400, and 800 at 1-$\sigma$,
2-$\sigma$, and 3-$\sigma$ levels, respectively, for advanced LIGO. If
the true equation of state is stiff and the typical neutron-star
radius is $R \agt 13$\,km, our analysis suggests that the radius will
be constrained within $\approx 1$\,km at 2-$\sigma$ level for an event
at $D_{\rm eff}=200$\,Mpc. On the other hand, if the true equation of
state is soft and the typical neutron-star radius is $R \alt 12$\,km,
it will be difficult to narrow down the equation of state among many
soft ones, although it is still possible to discriminate the true one
from stiff equations of state with $R \agt 13$\,km. We also find that
gravitational waves from binary neutron stars will be distinguished
from those from spinless binary black holes at more than 2-$\sigma$
level for an event at $D_{\rm eff}=200$\,Mpc. The validity of the EOB
formalism, Taylor-T4, and Taylor-F2 approximants as the inspiral
waveform model is also examined.

\end{abstract}


\maketitle

\section{Introduction}

The inspiral and merger of coalescing compact binaries are the most
promising sources for ground-based kilometer-size
laser-interferometric gravitational-wave
detectors~\cite{Detectors,aligo,avirgo,kagra}. Among them, the
advanced LIGO started the first observational run from September in
2015~\cite{aligo} and has achieved the first direct detection of
gravitational waves from the merger of a binary black
hole~\cite{0914}.  We may expect that these gravitational-wave
detectors will also detect the signals of gravitational waves from
binary-neutron-star mergers in a few years because the latest
statistical studies suggest that these gravitational-wave detectors
will observe gravitational waves from merger events as frequently as
$\sim 1$--$100$/yr if the designed sensitivity is
achieved~\cite{Kalogera,RateLIGO,Kim}.  One of the primary purposes
after the first detection of gravitational waves from binary neutron
stars (and also a black hole-neutron star binary) will be to extract
information of the neutron-star equation of state, which is still
poorly constrained~\cite{lattimer}. 

Extracting the {\em tidal deformability} of the neutron stars from
gravitational waves emitted by binary-neutron-star inspirals is one of
the most promising methods for constraining the neutron-star equation
of
state~\cite{lai94,mora04,flanagan08,tania10,VFT11,BDF12,read13,pozzo13,favata14,yagi14,wade14,agathos15}.
For this purpose, we need an accurate theoretical template of
gravitational waves from binary-neutron-star inspirals taking into
account tidal-deformation effects that influence the dynamics of the
late inspiral orbits and modify the corresponding gravitational
waveform.  However, current post-Newtonian (PN) waveforms are not
appropriate for the theoretical template as Favata~\cite{favata14} and
Yagi and Yunes~\cite{yagi14} independently showed that uncertainties in
the known PN waveforms can cause significant systematic errors in the
tidal deformability estimates due to the unknown higher-order terms.
In fact, Wade and his collaborators~\cite{wade14} evaluated the
systematic errors using the waveforms derived from different PN
families and confirmed that the estimated tidal deformability can be
significantly biased. To derive an accurate theoretical template that
is free from such uncertainties, high-accuracy numerical-relativity
simulations are necessary. Several efforts for this purpose have been
done
recently~\cite{thierfelder11,bernuzzi11,bernuzzi12,HKS2013,radice14,hotoke2015,bernuzzi15,bernuzzi14,haas}.

In our previous paper~\cite{hotoke2015}, we reported our latest effort
for deriving accurate gravitational waveforms from inspiraling binary
neutron stars of typical mass (1.35--1.35$M_\odot$). We performed
simulations for 15--16 inspiral orbits (30--32 wave cycles) up to the
merger employing low-eccentricity initial data, then performed an
extrapolation procedure with respect to the grid resolution, and
finally derived waveforms with the total accumulated phase error
within $\sim 0.5$ radian and amplitude error less than 2--3\%. We then
compared our numerical waveforms with the waveforms derived in an
effective-one-body (EOB) formalism, developed by Damour, Nagar, and
their collaborators~\cite{bernuzzi15} (see also
Refs.~\cite{DN2010,BDF12,damour12,DNB2013,BD2013,BD2014}). We have
indicated that the EOB results agree well with the
numerical-relativity results for a quite soft equation of state in
which the neutron-star radius is small ($\sim 11$\,km), while for a
stiff equation of state with the radius $\agt 13.5$\,km, a slight
disagreement is present for the final inspiral stage just prior to the
merger.

Combining numerical-relativity waveforms and resummed PN waveforms (by
a Taylor-T4 approximant), Read and her collaborators constructed
hybrid waveforms and analyzed the measurability of the tidal
deformability for the first time~\cite{read13}. The primary purpose of
our paper is to quantitatively update their previous results by
performing the same analysis as theirs using our new numerical
waveforms.  The motivation for this comes from the fact that the
quality of our numerical waveforms is significantly improved: (i) the
cycles of the new waveforms are double of those of the waveforms
previously used, (ii) the initial orbital eccentricity is reduced by
more than an order of magnitude~\cite{KST2014}, and (iii) the
convergence of the simulation results is much better and the numerical
error is much lower than the previous results.

As a first step for constructing hybrid waveforms, we will compare new
numerical gravitational waveforms for several equations of state
(different from those employed in our previous papers) with the EOB
waveforms, and will reconfirm the conclusion in our previous
paper~\cite{hotoke2015}.  Then, we will analyze the measurability of
the tidal deformability using the new hybrid waveforms constructed by
combining the numerical-relativity and EOB results.  In this paper, we
focus only on the measurability by ground-based advanced
gravitational-wave detectors. 

By comparing the hybrid waveforms derived from the numerical and EOB
results with them, we also examine the validity of other
analytic/semi-analytic methods for modeling gravitational waveforms,
paying special attention to Taylor-T4 (TT4) and Taylor-F2 (TF2)
approximants in which the tidal correction is incorporated up to the
first PN order (e.g., see Ref.~\cite{wade14}). We will indicate that
the current version of these Taylor approximants does not yield
waveforms as accurately as those by an EOB formalism for equal-mass
binary neutron stars, primarily because of the lack of the
higher-order PN terms.

The paper is organized as follows.  In Sec.~II, we briefly summarize
the formulation and numerical schemes employed in our
numerical-relativity study, and also list the equations of state
employed.  In Sec.~III, we present our new waveforms and compare them
with those by the EOB and TT4 approximants. We then construct hybrid
waveforms using the numerical and EOB waveforms.  In Sec.~IV, we
explore the measurability of the tidal deformability using the hybrid
waveforms.  We also assess the validity of the EOB, TT4 (hybrid-TT4),
and TF2 approximants for modeling inspiraling binary neutron stars.
Section~V is devoted to a summary.  Throughout this paper, we employ
the geometrical units of $c=G=1$ where $c$ and $G$ are the speed of
light and the gravitational constant, respectively.

\section{Deriving numerical waveforms}

We briefly summarize the formulation and numerical schemes of our
numerical-relativity simulation, equations of state employed, and a
method for deriving an extrapolated gravitational waveform from the
raw numerical-relativity results. 

\begin{table*}[t]
\centering
\caption{\label{tab1} Equations of state (EOS) employed, the maximum
  mass of spherical neutron stars for given EOS, circumferential
  radius, dimensionless tidal deformability, and tidal Love number of
  $l=(2,~3,~4)$ for spherical neutron stars of mass $1.35M_\odot$,
  angular velocity of initial data, location of the outer boundaries
  along each axis, and the finest grid spacing in the three different
  resolution runs. $m_0$ denotes the total mass of the system for the
  infinite orbital separation.  In this study, $m_0=2.7M_\odot$.  For
  $m_0 \Omega_0 \approx 0.0155$, the corresponding gravitational-wave
frequency is $\approx 371$\,Hz.}
\begin{tabular}{cccccccccc}
\hline\hline
~~EOS~~ &$M_{\rm max}\,(M_\odot)$ 
&~$R_{1.35}$\,(km)~&~~~$\Lambda$~~~&~~$k_{2,1.35}$~~ & ~~$k_{3,1.35}$~~ 
& ~~$k_{4,1.35}$~~&~~$m_0\Omega_0$~~ &$L$\,(km) 
& $\Delta x $\,(km)    \\
\hline
APR4&2.20& 11.09 & 322 & 0.0908 & 0.0234 & 0.00884 &~0.0156~&2572&~~0.167,~0.209,~0.251  \\
SFHo&2.06& 11.91 & 420 & 0.0829 & 0.0216 & 0.00766 &~0.0155~&2858&~~0.155,~0.186,~0.233  \\
DD2 &2.42& 13.20 & 854 & 0.1007 & 0.0272 & 0.00996 &~0.0155~&3258&~~0.177,~0.212,~0.265  \\
TMA &2.02& 13.85 &1192 & 0.1103 & 0.0316 & 0.01229 &~0.0155~&3430&~~0.186,~0.223,~0.279  \\
TM1 &2.21& 14.48 &1428 & 0.1059 & 0.0300 & 0.01154 &~0.0155~&3644&~~0.198,~0.237,~0.297  \\
\hline\hline
\end{tabular}
\end{table*} 

\subsection{Evolution and initial condition}

We follow the inspiral, merger, and early stage of the post-merger of
binary neutron stars using our numerical-relativity code, {\tt SACRA},
for which the details are described in Ref.~\cite{yamamoto08}. As in
our previous long-term simulations~\cite{hotoke2015}, we employ a
moving puncture version of the Baumgarte-Shapiro-Shibata-Nakamura
formalism~\cite{BSSN}, {\em locally} incorporating a Z4c-type
constraint propagation prescription~\cite{Z4c} (see
Ref.~\cite{KST2014} for our implementation) for a solution of
Einstein's equation.  {\tt SACRA} implements a fourth-order finite
differencing scheme in space and time with an adaptive mesh refinement
(AMR) algorithm.

As in Ref.~\cite{hotoke2015}, we prepare nine refinement levels and
thirteen domains for the AMR algorithm. Each refinement domain
consists of a uniform, vertex-centered Cartesian grid with
$(2N+1,2N+1,N+1)$ grid points for $(x,y,z)$; the equatorial plane
symmetry at $z=0$ is imposed. The half of the edge length of the
largest domain (i.e., the distance from the origin to outer boundaries
along each axis) is denoted by $L$, which is chosen to be larger than
$\lambda_{0}$, where $\lambda_{0} = \pi/\Omega_{0}$ is the initial
wavelength of gravitational waves and $\Omega_{0}$ is the initial
orbital angular velocity. The grid spacing for each refinement level
is $\Delta x_{l}=L/(2^{l}N)$ where $l=0-8$. We denote $\Delta x_8$ by
$\Delta x$ in the following.  In this work, we choose $N=72$, 60, and
48 for examining the convergence properties of numerical results with
respect to the grid resolution. With the highest grid resolution (for
$N=72$), the semimajor diameter of each neutron star is covered by
about 120 grid points.

We prepare binary neutron stars in quasi-circular orbits for the
initial condition of numerical simulations. The initial conditions are
numerically obtained by using a spectral-method library,
LORENE~\cite{lorene}. In this paper, we focus only on equal-mass
systems with each neutron-star mass $1.35M_\odot$. We follow 14--16
orbits in this study ($\approx 57$--62\,ms duration for the last
inspiral orbits). To do so, the orbital angular velocity of the
initial configuration is chosen to be $m_0\Omega_{0} \approx 0.0155$
($f=\Omega_0/\pi \approx 371$\,Hz for the total mass
$m_0=2.7M_{\odot}$ where $f$ denotes the gravitational-wave
frequency).  Some of parameters for the models and setting for the
simulations are listed in Table~\ref{tab1}.

For the computation of an accurate gravitational waveform in numerical
simulations, we have to employ initial data of a quasi-circular orbit
of negligible eccentricity. Such initial data are constructed by an
eccentricity-reduction procedure described in Ref.~\cite{KST2014}.
For the initial data employed in this work, the residual eccentricity
is $\alt 10^{-3}$.

\subsection{Equation of state}

We employ four tabulated equations of state for zero-temperature
neutron-star matter derived recently by Hempel and his collaborators,
and we refer to them as SFHo~\cite{SFHo}, DD2~\cite{DD2},
TMA~\cite{TM1}, and TM1~\cite{TM1}.  Here, TM1 employed the same
parameter set of a relativistic mean-field theory as that of one of
Shen equations of state~\cite{Shen}.  All these equations of state
have been derived in relativistic mean field theories.  Some
characteristic properties resulting from these equations of state are
listed in Table~\ref{tab1}. For all of them, the predicted maximum
mass for spherical neutron stars is larger than the largest
well-measured mass of neutron stars, $\approx
2M_\odot$~\cite{demorest10}.  The neutron-star radius with mass
$1.35M_\odot$, $R_{1.35}$, is $\approx 11.9$, 13.2, 13.9 and 14.5\,km
for SFHo, DD2, TMA, and TM1; i.e., these are soft, moderately stiff,
stiff, and very stiff equations of state, respectively.

In our previous works~\cite{HKS2013,hotoke2015}, we employed
piece-wise polytropic equations of state approximating tabulated
equations of state. In this work, we employ the tabulated equations of
state as it is for preserving the original form of each equation of
state.

In the analysis for the measurability of the dimensionless tidal
deformability, $\Lambda$, we also employ the numerical results for
APR4~\cite{APR4}, for which a detailed numerical result has been
already reported in Ref.~\cite{hotoke2015}.  For this numerical
simulation, we employed the piece-wise polytropic
approximation. $R_{1.35}$ for this equations of state is $\approx
11.1$\,km, and hence, this equation of state is softer than SFHo, DD2,
TMA, and TM1. As Table~\ref{tab1} shows, $R_{1.35}$ and $\Lambda$ are
systematically varied among the five equations of state employed.
This is the reason that we pick up these equations of state in our
present analysis for the measurability of $\Lambda$.

For the zero-temperature case, the thermodynamical quantities, i.e.,
the pressure, $P$, and the specific internal energy, $\varep$, are
written as functions of the rest-mass density, $\rho$. Here, the
zero-temperature equations of state satisfy $d\varep=-Pd(1/\rho)$.  In
numerical simulations, we slightly modify the original equations of
state, adding a thermal part, to approximately take into account
thermal effects, which play a role in the merger and post-merger
phases.  For this prescription, we use the same method as that used in
our previous works (see, e.g.,
Refs.~\cite{Hotokezaka2013,hotoke2015}).

\subsection{Extraction of gravitational waves and extrapolation procedures}

Gravitational waves are extracted from the outgoing-component of
complex Weyl scalar $\Psi_{4}$~\cite{yamamoto08}.  $\Psi_{4}$ can be
expanded in the form
\begin{eqnarray}
\Psi_{4}(t, r, \theta, \varphi) = \sum_{lm}\Psi_{4}^{l,m}(t,r)_{-2}
Y_{lm}(\theta, \varphi),
\end{eqnarray}
where $_{-2}Y_{lm}(\theta, \varphi)$ denotes the spin-weighted
spherical harmonics of weight $-2$ and $\Psi_{4}^{l,m}$ are expansion
coefficients defined by this equation.  In this work, we focus only on
the $(l,|m|)=(2,2)$ mode because we pay attention only to the
equal-mass binary, and hence, this quadrupole mode is the dominant
one.

From the $(l,m)=(2,2)$ mode, quadrupole gravitational waveforms are
determined by
\begin{eqnarray}
h_{+}(t,r)-i h_{\times}(t,r) = -\lim_{r\rightarrow \infty}
\int^{t}dt^{\prime}\int^{t^{\prime}}dt^{\prime \prime}
\Psi_{4}^{2,2}(t^{\prime \prime},r),\nonumber \\ \label{eq:hpsi4}
\end{eqnarray}
where $h_+(t,r)$ and $h_\times(t,r)$ are the plus and cross modes of
quadrupole gravitational waves, respectively (note that the waveforms
$h_+$ and $h_\times$ are actually derived by the integration method of
Ref.~\cite{RP2011}: see also Ref.~\cite{KST2014,hotoke2015}).

We evaluate $\Psi_{4}$ at a finite spherical-coordinate radius, $r
\approx 200m_0$, following Ref.~\cite{hotoke2015}.  The waveforms are
described as a function of the retarded time defined by
\begin{eqnarray}
t_{\rm{ret}} := t - r_{*}, \label{eq:tret1}
\end{eqnarray}
where $r_{*}$ is the so-called tortoise coordinate defined by
\begin{eqnarray}
r_{*} := r_{\rm{A}} + 2m_0\ln \left(\frac{r_{\rm{A}}}{2m_0}-1\right),
 \label{eq:tret2}
\end{eqnarray}
with $r_{\rm{A}}:=\sqrt{A/4\pi}$ and $A$ the proper area of the
extraction sphere: For simplicity we define it by $r_{\rm
  A}=r[1+m_0/(2r)]^2$.

Since the waveform of $\Psi_4^{2,2}$ extracted at a finite radius,
$r=r_0$, is systematically different from that at null infinity, we then
compute an extrapolated waveform at $r_0 \rightarrow \infty$ using the
Nakano's method as~\cite{LNZC2010,Nakano15}
\beqn
\Psi_4^{l,m,\infty}(t_{\rm ret}, r_0)
&=&C(r_0)\biggl[\Psi_4^{l,m}(t_{\rm ret}, r_0) \nonumber \\
&~& -{(l-1)(l+2) \over 2r_{\rm A}}
\int^{t_{\rm ret}} \Psi_4^{l,m}(t', r_0)dt' \biggr], \nonumber \\
\label{eq:nakano}
\eeqn
where $C(r_0)=1 - 2m_0/r_{\rm A}$ as described in Ref.~\cite{hotoke2015}. 

As we already mentioned, we always perform simulations for three
different grid resolutions (with different values of the grid spacing
$\Delta x$), and obtain three waveforms of different accuracy
determined by $\Delta x$.  Then, we perform an extrapolation procedure
for $\Delta x \rightarrow 0$ employing the same method as described in
Ref.~\cite{hotoke2015}. As in the previous results, we found that the
convergence order is within $4 \pm 1$ irrespective of the equations of
state employed.

It should be noted that the extrapolated numerical waveforms have the
accumulated phase errors only within $\sim 0.5$ radian as described 
in Ref.~\cite{hotoke2015}.  This value is much smaller than the phase
differences among the different waveforms by different modeling~(see
Sec.~IV B). Therefore, we expect that the numerical errors in our
extrapolated waveforms do not change significantly our results for the
analysis of the measurability described in Sec.~IV. 

\section{Constructing hybrid waveform}

\begin{figure*}[t]
\begin{center}
\includegraphics[width=86mm]{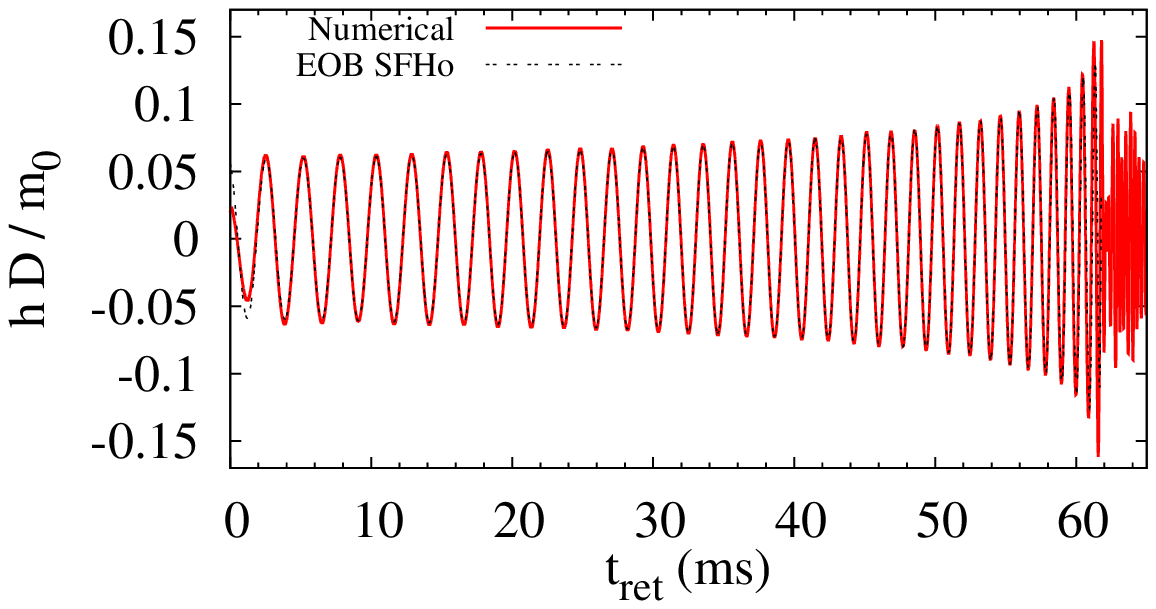}
~~~\includegraphics[width=86mm]{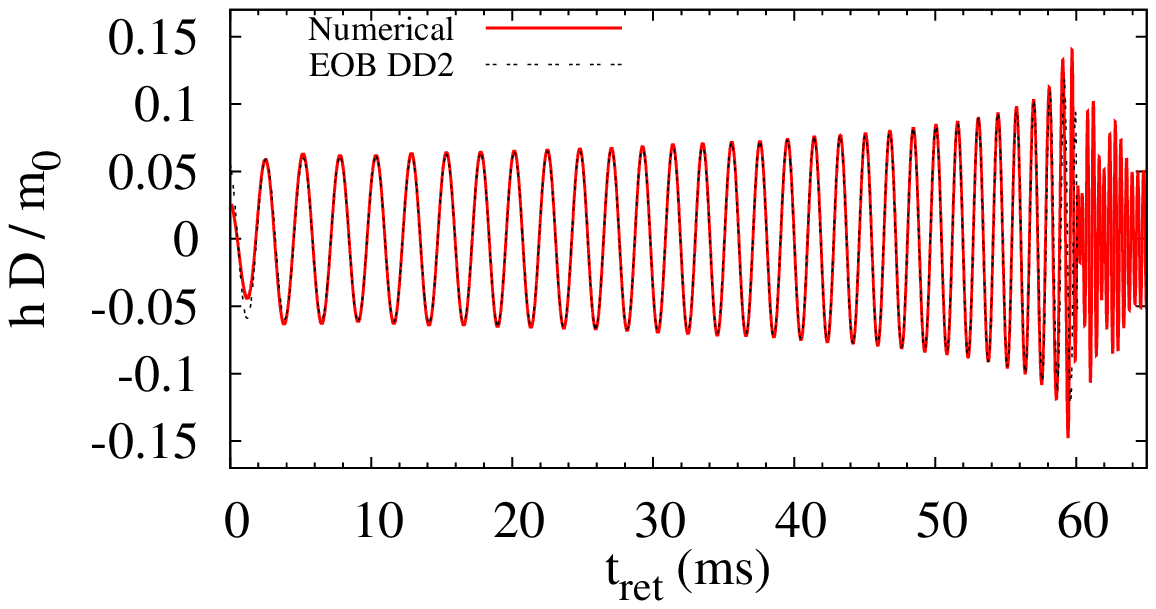}\\
\includegraphics[width=86mm]{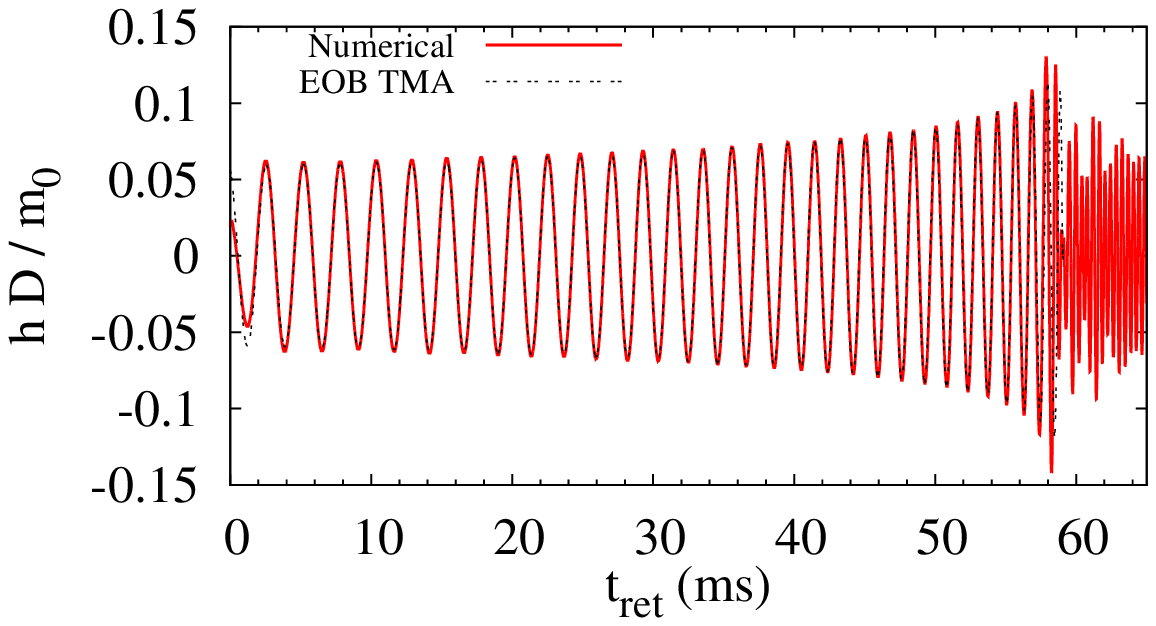}
~~~\includegraphics[width=86mm]{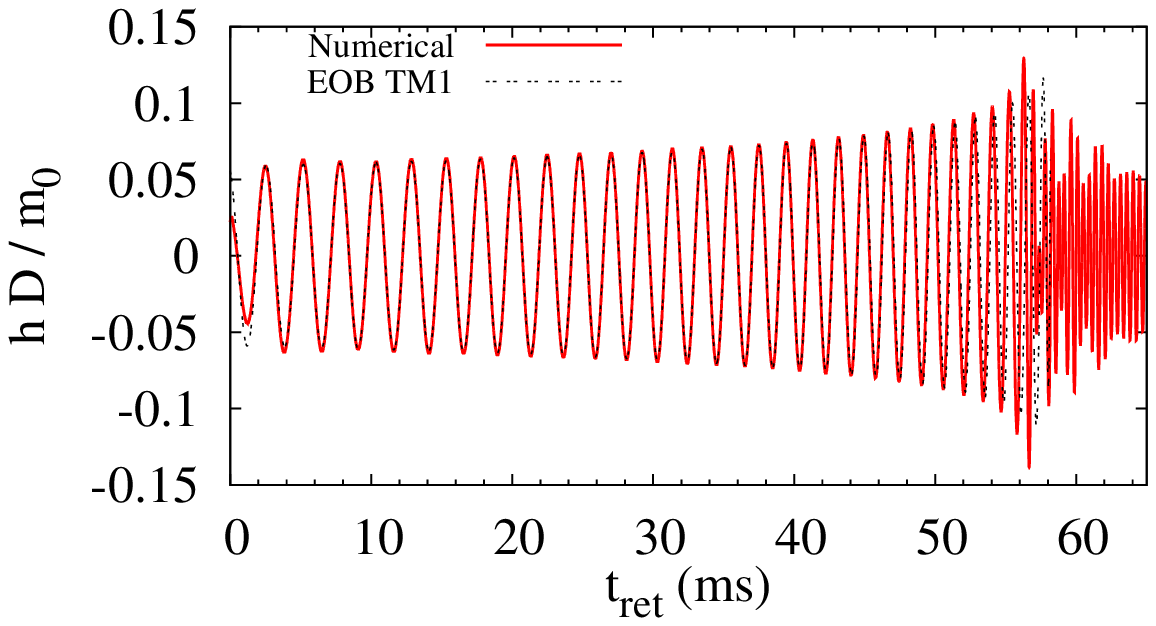}
\caption{Comparison of numerical (solid curves) and EOB (dot-dot
  curves) waveforms for the late inspiral phase. Upper left, upper
  right, lower left, and lower right panels show the results for SFHo,
  DD2, TMA, and TM1, respectively. Gravitational waves (plus mode)
  observed along the rotational axis (perpendicular to the orbital
  plane) are shown.  $D$ denotes the distance from the source to the
  observer.
\label{fig1}}
\end{center}
\end{figure*}

\begin{figure*}[t]
\begin{center}
\includegraphics[width=86mm]{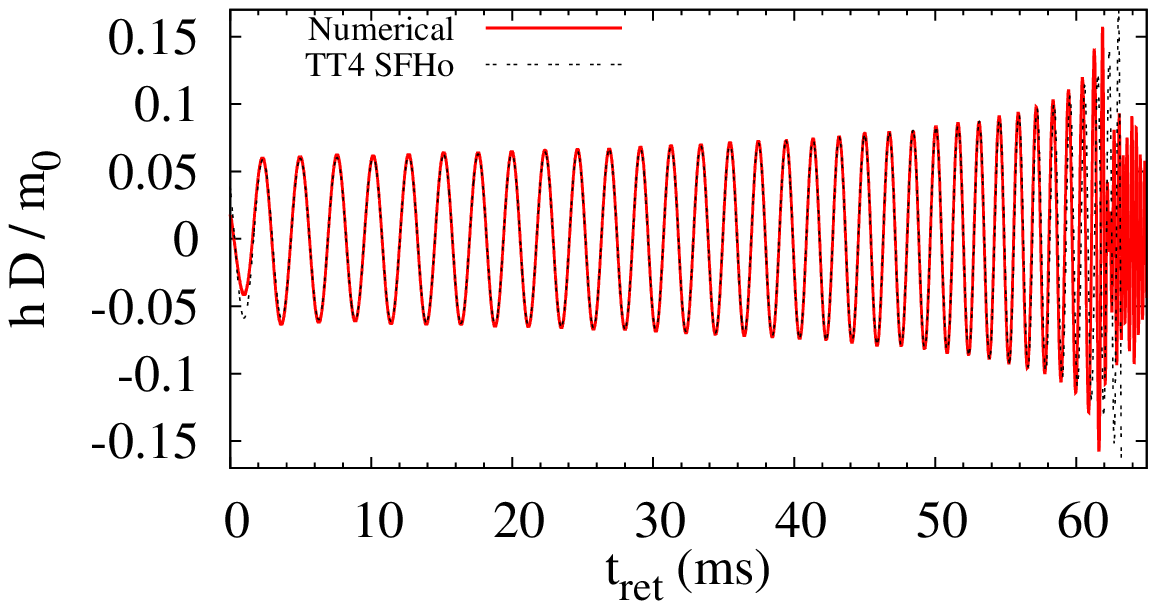}
~~~\includegraphics[width=86mm]{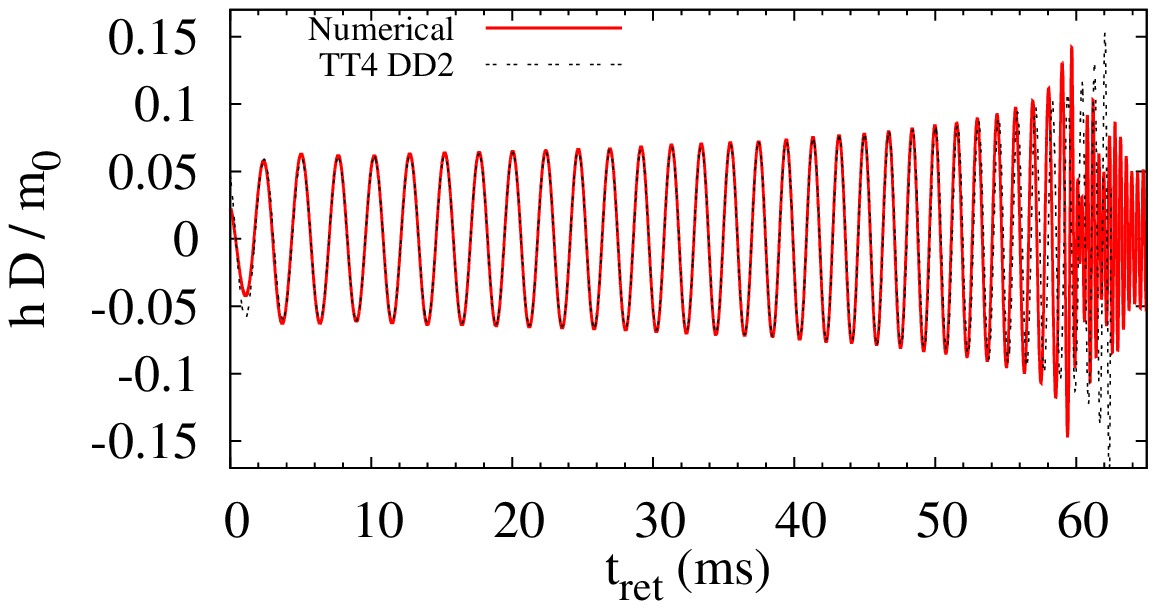}\\
\vspace{0.2cm}
\includegraphics[width=86mm]{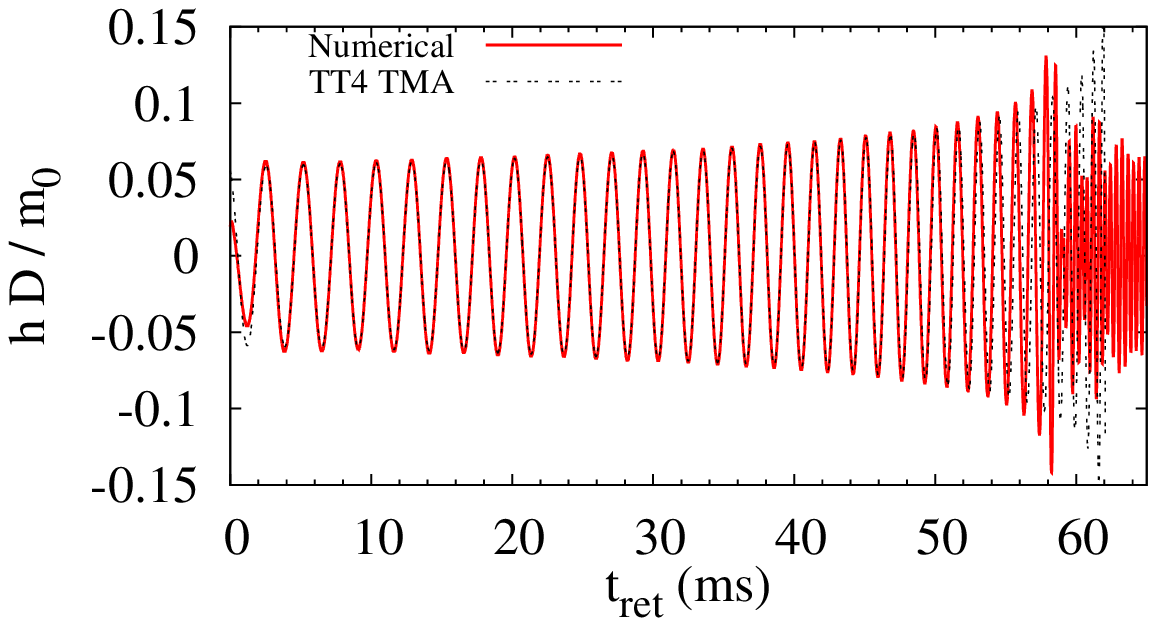}
~~~\includegraphics[width=86mm]{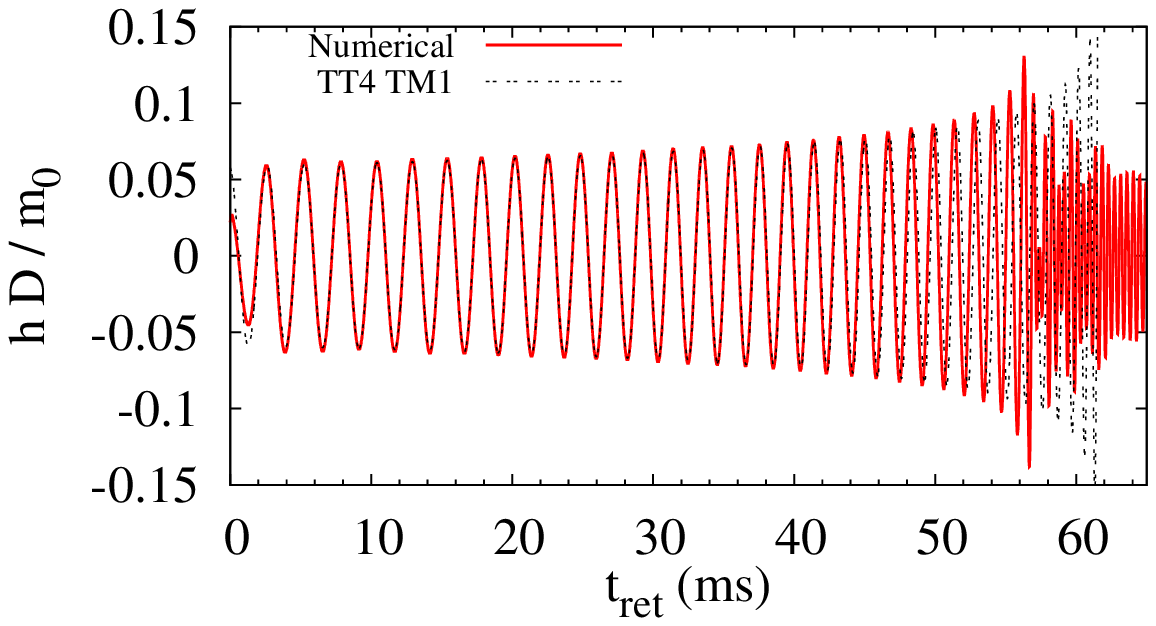}
\caption{The same as Fig.~\ref{fig1}, but for the case that TT4
waveforms are used for the comparison with the numerical waveforms.
\label{fig2}}
\end{center}
\end{figure*}

Because we follow only 14--16 inspiral orbits, gravitational waveforms
only with $f \agt 370$\,Hz can be derived. For exploring the
measurability of the tidal deformability, such waveforms are not
well-suited. To supplement the earlier waveform for $f < 370$\,Hz, we
consider hybridization between the numerical waveform and a waveform
derived by an analytic/semi-analytic calculation, by which the
waveform for the lower-frequency band is filled up.

For the hybridization, we first have to align the time and phase of
the numerical-relativity waveform, $h_{\rm NR}$, and a waveform by an
analytic/semi-analytic formulation, $h_{\rm SA}$.  Here, complex
waveforms $h(t)$ are defined by $h_+(t) - i h_\times(t)$ with $h_+(t)$
and $h_\times(t)$ the plus and cross modes, respectively. We then
calculate
\beqn
I(\tau,\phi)&=&\int_{t_i}^{t_f} dt 
\left|h_{\rm NR}(t)-h_{\rm SA}(t+\tau)e^{i\phi}\right|^2, 
\label{corre}
\eeqn
and search for $\tau$ and $\phi$ that minimize $I$.  Here, $t$ in this
section always denotes the retarded time, $t_{\rm ret}$, and we choose
$t_i=5$\,ms and $t_f=20$\,ms as in our previous
paper~\cite{hotoke2015}. At $t=5$\,ms and 20\,ms, the
gravitational-wave frequency is $f \approx 380$\,Hz and 420\,Hz,
respectively (see Fig.~\ref{figap0} in Appendix~A), and the number of
the wave cycle in this duration is $\sim 6$ (see Fig.~1). We choose
this window because we would like to employ the time for it as early
as possible.  Here, for the first $\approx 5$\,ms just after the
simulations started, the waveforms have unphysical modulation, and
hence, we choose 5\,ms for $t_i$.  20\,ms for $t_f$ is rather ad hoc.
To check that our conclusion for the measurability of the tidal
deformability does not depend strongly on the choice of $t_i$ and
$t_f$, we also construct another hybrid waveforms choosing
$t_i=10$\,ms and $t_f=25$\,ms ($f \approx 390$\,Hz and 430\,Hz,
respectively) and use them for calibrating the results in Sec.~IV A.

For the values of $\tau$ and $\phi$ that we determine, we construct a
hybrid waveform. Following Refs.~\cite{hotoke2015,lackey14}, we define
the hybrid waveform by
\beqn
&&h_{\rm hyb}(t) \nonumber \\
&&=
\left\{
\begin{array}{lc}
h_{\rm SA}(t')e^{i\phi} & t \leq t_i, \\
h_{\rm NR}(t) H(t) + h_{\rm SA}(t')e^{i\phi}[1-H(t)] &~~ t_i \leq t \leq t_f, \\
h_{\rm NR}(t) & t \geq t_f,
\end{array}
\right.\nonumber \\
\eeqn
where $t'=t+\tau$, and we choose a Hann window function for $H(t)$ as
\beqn
H(t):={1 \over 2}\left[1 - \cos\left(\pi{t-t_i \over t_f-t_i}\right)\right]. 
\eeqn

Figure~\ref{fig1} plots the extrapolated numerical waveforms and
waveforms by an EOB formalism~\cite{bernuzzi14} for four different
equations of state (see Fig.~3 and Appendix~A of
Ref.~\cite{hotoke2015} for the waveform with APR4 and for the EOB
formalism that we employ in this work, respectively). For these plots,
we align the numerical and EOB waveforms in the same way as the hybrid
construction.  It is found that the two waveforms agree well with each
other in their early part, i.e., for $t_{\rm ret} \alt 45$\,ms (for
the first $\sim 20$ wave cycles). In particular, the phases for the
two waveforms agree with each other with the disagreement of order
0.01\,rad for this stage, as we demonstrated in our previous
work~\cite{hotoke2015}. This suggests that our hybridization would
work well whenever we employ the EOB waveforms irrespective of the
choice of $(t_i, t_f)$ as long as they are sufficiently small $\ll
45$\,ms.  We estimate a degree of the disagreement in the matching
window by
\beqn
\left[{\min_{\tau , \phi} I(\tau, \phi) \over 
{\int_{t_i}^{t_f} dt \left\{ \left|h_{\rm NR}(t)\right|^2
+\left|h_{\rm SA}(t)\right|^2\right\}}} \right]^{1/2}, 
\label{corre2}
\eeqn
and it is always small as $\alt 2\times 10^{-2}$.  This error comes
primarily from the error in amplitude of the numerical waveforms
because the estimated maximum error size is 2--3\% in the amplitude.
On the other hand, the phase error has a minor contribution for this
error.

The numerical and EOB waveforms agree reasonably well with each other
even in the late inspiral phase, up to a few wave cycles prior to the
merger (see also Fig.~\ref{figap0} in Appendix A for supplementary
information).  This indicates that the tidal-deformation effects would
be fairly well taken into account in the employed EOB formalism as we
already mentioned in Ref.~\cite{hotoke2015}.  Due to these reasons, we
construct hybrid waveforms employing the EOB waveforms as $h_{\rm SA}$
and use them for analyzing the measurability of the tidal
deformability.

We note that for stiff equations of state like TMA and TM1 for which
the dimensionless tidal deformability is larger than $1000$, the
disagreement between the numerical and EOB waveforms are appreciable
for the last few wave cycles, as already pointed out in
Ref.~\cite{hotoke2015}. This suggests that there is still a room for
incorporating additional tidal effects into the EOB
formalism~\cite{tania}.  On the other hand, for softer equations of
state with $\Lambda < 1000$, the disagreement is minor. This indicates
that the EOB waveforms well capture the tidal-deformation effects as
long as $\Lambda \ll 1000$. 

\begin{figure*}[t]
\begin{center}
\includegraphics[width=110mm]{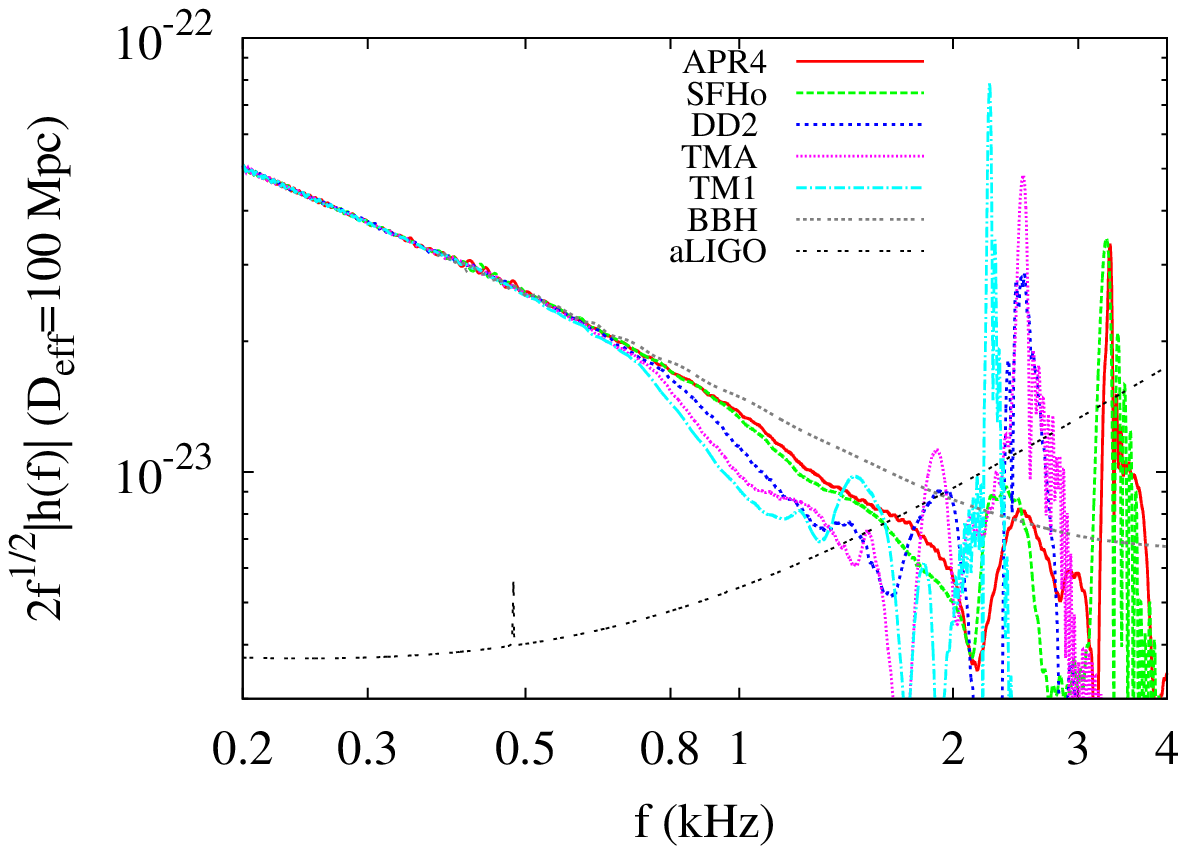}
\caption{Fourier spectra of the hybrid waveforms for five different
  equations of state for a hypothetical effective distance of $D_{\rm
    eff}=100$\,Mpc. The dot-dot curve for the advanced LIGO (referrer
  to as aLIGO) denotes $S_n^{1/2}$. Here, $S_n$ is the one-sided noise
  spectrum density for the ``Zero Detuning High Power''
  configuration~\cite{ligonoise}. The dot-dot-dot curve denotes the
  Fourier spectrum for a spinless binary black hole of mass
  1.35--1.35$M_\odot$ (plotted only for $f \geq 375$\,Hz). To
  approximately find SNR, the spectrum is shown with the additional
  factor of 2: see Eq.~(\ref{eqSNR}).
\label{fig3}}
\end{center}
\end{figure*}


We also perform the hybridization employing the TT4
waveforms~\cite{Boyle07,Ajith07} incorporating the tidal effects up to
the first PN (1PN) order~\cite{VFT11}. In the TT4 approximant, the
evolution of the gravitational-wave frequency is determined by (see,
e.g., Ref.~\cite{wade14})
\beqn
{dx \over dt}
&=&{16 \over 5 m_0} x^{5}
\biggl[1 -{487 \over 168}x +4\pi x^{3/2}
\nonumber \\
&&~~~~+{274 229 \over 72 576} x^2
-{254 \over 21}\pi x^{5/2} 
\nonumber \\
&&~~~~+\biggl( {178 384 023 737 \over 3 353 011 200} + {1475\pi^2 \over 192}
-{1712 \over 105}\gamma_{\rm E} \nonumber \\
&&~~~~~~~-{856 \over 105}\ln (16x) \biggr)x^3
+ {3310 \over 189} \pi x^{7/2} \nonumber \\
&&~~~~+{39 \over 8}\Lambda x^5 + {5203 \over 896}\Lambda x^6 
\biggr], \label{eq:TT4}
\eeqn
where $x(t):=[\pi m_0 f(t)]^{2/3}$ and $\gamma_{\rm E}$ is the Euler's 
constant. 
We assume that the quadrupole-wave amplitude is determined by Eq.~(71)
of Ref.~\cite{Boyle08}.
Here, for simplicity, we restrict our attention only to the equal-mass
case, and in addition, we do not take into account the effect of the tidal
deformability in the amplitude because it plays only a minor role
for analyzing the measurability~\cite{damour12}. 

After the alignment procedure for time and phase, we also compare the
numerical waveforms with the TT4 waveforms. Figure~\ref{fig2} shows
the results for the same comparison as in Fig.~\ref{fig1}.  This shows
that the agreement between the numerical and TT4 waveforms is worse than
that between the numerical and EOB waveforms. Specifically, the phase
evolution in the TT4 approximant is slower than that in the EOB
formalism.  We note that the tidal effects accelerate the orbital
evolution in the late inspiral phase because the tidal force
strengthens the attractive force between two neutron stars for such
orbits.  Thus, we conclude that the tidal effects are underestimated
in the employed TT4 approximant. This should be the case not only for
the very late inspiral phase but also for the earlier inspiral
phase. An analysis of the gravitational-wave phase evolution indicates
that this would be due to the lack of the higher-order PN terms of order
$O(x^{13/2})$ or more: terms with more than 1.5PN order with respect
to the leading-order tidal-deformation effect.  We indicate evidence
for this in Appendix~B.  By this reason, we suppose that the EOB
formalism could give a better waveform than the TT4 formalism. 



Figure~\ref{fig3} plots the Fourier spectra of the hybrid waveforms
(numerical plus EOB waveforms) together with a designed noise curve of
the advanced LIGO, $S_n^{1/2}$ (for ``Zero Detuning High Power''
configuration)~\cite{ligonoise} and with the spectrum of a
binary-black-hole merger of mass 1.35--1.35$M_\odot$.  Here, $S_n(f)$
denotes the one-sided noise spectrum density of gravitational-wave
detectors.  The numerical waveform for the binary black hole is taken
from SXS Gravitational Waveform Database~\cite{SXS} and we employ
SXS:BBH:001. In this paper, the Fourier transform is defined by
\beqn
\tilde{h}(f):= \int dt \,h_+(t) \exp(-2\pi i ft), 
\eeqn
where $h_+(t)$ denotes the plus-mode gravitational waveform. For
binary neutron stars, the overall shape of $h_\times(t)$ is
approximately the same as that of $h_+(t)$ except for a $\pi/2$ phase
difference, and hence, the Fourier transformation of the cross mode,
$h_\times(t)$, results approximately in $-i \tilde{h}(f)$. 

The response of gravitational-wave detectors for a gravitational-wave
event of coalescing binary neutron stars is written in the form
\beqn
\bar{h}(t)=H_+(\theta, \varphi, \iota, \psi_p)h_+(t)
+H_\times(\theta, \varphi, \iota, \psi_p)h_\times(t),
\nonumber \\
\eeqn
where $H_+$ and $H_\times$ are functions of the source angular
direction denoted by $(\theta, \varphi)$, of the inclination
angle of the binary orbital plane with respect to the line of the
sight to the source denoted by $\iota$, and of the 
polarization angle denoted by $\psi_p$.  Thus, the Fourier
transformation of $\bar{h}(t)$ is written as
\beqn
\bar{h}(f) \approx 
H(\theta,\varphi,\iota,\psi_p)\tilde{h}(f), 
\eeqn
where $H=H_+ -i H_\times$ for which $|H|\leq 1$. Taking into account
this form, we define the effective distance to the source by $D_{\rm
  eff}:=D |H|^{-1}$ where $D$ is the proper distance to the source. In
the following, we always refer to $D_{\rm eff}$ (not to $D$) as ``the
effective distance to the source'', and we typically consider an event
at $D_{\rm eff}=200$\,Mpc: This is equivalent of an event at a
distance of 200\,Mpc with the optimal orientation and sky
location. The reason for this choice is that statistical studies have
predicted typically $\sim 1$ detection per year for $D_{\rm eff} \alt
200$\,Mpc~\cite{abadie10}.


Figure~\ref{fig3} clearly shows that the difference in the Fourier
spectra among the waveforms of different equations of state becomes
appreciable for $f \agt 500$\,Hz. In particular, for $f \agt 700\,{\rm
  Hz}$, the difference is remarkable. This stems primarily from the
difference in the tidal deformability: For the larger values of
$\Lambda$, the spectrum amplitude more steeply decreases for $f \agt
700$\,Hz because the binary orbit is evolved faster.  
Here, we note that (i) the late inspiral waveform determines the
spectrum only for $f \alt 1$\,kHz, (ii) the final-inspiral to merger
waveform determines the spectrum approximately for 1\,kHz$\alt f \alt
2$\,kHz, and (iii) several bumps and peaks for $f \agt 2$\,kHz are
determined by the post-merger waveform (i.e., by gravitational waves
from remnant massive neutron stars formed after the merger).  It
should be also noted that the noise amplitude of the
gravitational-wave detectors monotonically increases for $f \agt
500$\,Hz. This indicates that the equation of state (tidal
deformability) could be constrained primarily by analyzing the
spectrum in the late inspiral and merger waveforms and that the tidal
deformability could be more accurately measured for stiffer equations
of state of a larger value of $\Lambda$.

We briefly comment on the strength of gravitational waves found for
2--3.5\,kHz as peaks of the Fourier spectrum, which are emitted from
the massive neutron stars formed in the post-merger phase (see, e.g.,
Ref.~\cite{Hotokezaka2013}). To assess the {\em detectability} for
them, we estimate the signal-to-noise ratio defined by
\begin{equation}
{\rm SNR} = \left[ 4 \int_{f_i}^{f_f} {|\tilde{h}(f)|^2 \over S_n(f)}
df \right]^{1/2}. \label{eqSNR}
\end{equation}
For evaluating the strength of the peaks, we choose $f_i=2$\,kHz and
$f_f=4$\,kHz and the one-sided noise spectrum density for the ``Zero
Detuning High Power'' configuration of advanced LIGO as $S_n
(f)$~\cite{ligonoise}.  It is found that SNR is 0.5--0.9 for $D_{\rm
  eff}=200$\,Mpc: For stiffer equations of state, this value is larger
(0.5, 0.6, 0.7, 0.9, and 0.9 for APR4, SFHo, DD2, TMA, and TM1).
Since SNR$\agt 5$ would be required for the confirmed detection (due
to the presence of the Gaussian and other noises in the detectors;
see, e.g.,~Ref.~\cite{clark14}), this peak will be detected with a
high confidence level only for a nearby event with $D_{\rm eff}\alt
20$--35\,Mpc for the advanced-LIGO-class detectors, even if perfect
templates for this waveform could be prepared.  We note that for a
gravitational-wave event of equal-mass binary neutron star with
$m_0=2.7M_\odot$ and $D_{\rm eff}=200$\,Mpc, the total signal-to-noise
ratio for the entire inspiral phase will be $\approx 17$ (for a choice
of $f_i<10$\,Hz and $f_f > 2$\,kHz) irrespective of the equations of
state employed.  Therefore, the expected SNR for the kHz-peaks is much
lower than the SNR for the inspiral signal for the advanced-LIGO-class
detectors.  This motivates us to focus primarily on the late inspiral
phase for extracting the information of the neutron-star equation of
state at least in the near future. (Of course, detectability of this
peak will be more optimistic with more sensitive gravitational-wave
detectors in the future.)

\section{Measurability of the tidal deformability}

Following Ref.~\cite{read13}, we define a measure of the distinguishability 
of two waveforms by 
\beqn
&&|| h_1 - h_2 ||^2  \nonumber \\
&&:= \min_{\Delta t , \Delta \phi} \left[ 4 \int_{f_i}^{f_f}
    \frac{\left|\tilde{h}_1(f) - \tilde{h}_2(f)e^{i(2\pi f \Delta t + \Delta \phi)} 
 \right|^2}{S_n (f)} 
df \right], \nonumber \\
\label{dSNR}
\eeqn
where $\tilde{h}_1(f)$ and $\tilde{h}_2(f)$ are the Fourier transform
of the waveforms $h_1(t)$ and $h_2(t)$.  $f_i$ and $f_f$ are carefully
chosen later for the analysis of the measurability.  In the following,
we always employ the one-sided noise spectrum density for the ``Zero
Detuning High Power'' configuration of advanced LIGO as $S_n
(f)$~\cite{ligonoise}.

As shown in Ref.~\cite{lee08}, $||h_1-h_2||=1$ corresponds to a
1-$\sigma$ error in parameter estimation, and hence, two waveforms
$h_1$ and $h_2$ are said to be marginally distinguishable if
$||h_1-h_2||=1$. Thus, we assess the measurability of the tidal
deformability by calculating $||h_1-h_2||$ for a variety of waveform
combinations. 

In the calculation of $||h_1-h_2||$, it is ideal to choose
$f_i<10$\,Hz and $f_f > 4$\,kHz.  Computationally, choosing $f_f >
4$\,kHz does not matter whereas choosing the low value of $f_i$ is
expensive because the data size for the waveforms increases
approximately as $f_i^{-8/3}$.  Here, we should keep in mind that the
noise amplitude of ground-based gravitational-wave detectors steeply
increases with the decrease of the frequency for $f < 50$\,Hz toward
10\,Hz. Hence, it is practically possible to obtain an approximate
result for $||h_1-h_2||$ even if we choose a value of $f_i$ that is
larger than 10\,Hz.  Thus, as a first step, we calibrated how high
value of $f_i$ would be acceptable analyzing $||h_1-h_2||$ by using
a TF2 approximant for $\tilde{h}_1$ and $\tilde{h}_2$. Here, the
amplitude and phase of the TF2 approximant are calculated by using a
stationary phase approximation and the results are written simply in a
polynomial form with respect to $(\pi m_0 f)^{2/3}$~\cite{Ajith07}
(see also Appendix~C).  In the present analysis, the tidal effect is
incorporated up to the 1PN order as in the TT4 case.

It is found (see Appendix~C for the results) that for $f_i=30$ and
50\,Hz, the results for $||h_1-h_2||$ are not significantly different
from that for $f_i=10$\,Hz: the values of $||h_1-h_2||$ are
systematically underestimated by $\approx 5$\% and 15\% for $f_i=30$ and
50\,Hz, respectively: see Appendix~C. For $f_i=100$\,Hz, the values of
$||h_1-h_2||$ are underestimated by up to $\sim 30$\%. (We note that
for $f_i=10$, 30, 50, and 100\,Hz with $f_f=4$\,kHz, SNR of
Eq.~(\ref{eqSNR}) is $\approx 17$, 16, 13, and 9 for $D_{\rm
  eff}=200$\,Mpc.)  Thus, in this paper, we employ $f_i=30$\,Hz for the
analysis of the measurability of the tidal deformability performed in
Sec.~IV A and $f_i= 50$\,Hz for the calibration of the several
model waveforms (see Sec.~IV B).

For $f_f$, we choose 1\,kHz, 2\,kHz, and 4\,kHz.  As we already
mentioned, the contribution to the SNR from $f \geq 2\,{\rm kHz}$ is 
minor, and hence, the results for $||h_1-h_2||$ with $f_i=2$\,kHz and
4\,kHz are approximately identical (see Sec.~IV A and Appendix~C).

\begin{table*}[t]
\caption{$|| h_1 - h_2 ||$ for combination of hybrid waveforms with
  different equations of state for an event of $D_{\rm eff}=200$\,Mpc
  for which the total SNR would be $\approx 17$ for our choice of
  $S_n$ (for $f_i \alt 10$\,Hz and $f_f \agt 4$\,kHz).  The left and
  right tables show the results for $f_i=30$ and 50\,Hz, respectively.
  For the top, second, and third tables, $f_f=1$, 2, and 4\,kHz,
  respectively.  Note that the listed values are proportional to
  $200\,{\rm Mpc}/D_{\rm eff}$ and for $f_i=30$\,Hz and 50\,Hz, the
  values for given combination of two waveforms would be smaller than
  those for $f_i=10$\,Hz by $\approx 5$\% and 15\%, respectively (see
  Appendix~C).
\label{table2}}
 \begin{tabular}{c|ccccc}
  0.03--1\,kHz& APR4 & SFHo & DD2 & TMA & TM1 \\
  \hline
  APR4 & ---  & 0.4  & 2.2 & 2.9 & 3.4 \\
  SFHo & 0.4  & ---  & 1.9 & 2.7 & 3.2 \\
  DD2  & 2.2  & 1.9  & --- & 1.3 & 2.4 \\
  TMA  & 3.2  & 2.7  & 1.3 & --- & 1.6 \\
  TM1  & 3.4  & 3.2  & 2.4 & 1.6 & --- 
 \end{tabular}
\hspace{1cm}
 \begin{tabular}{c|ccccc}
  0.05--1\,kHz& APR4 & SFHo & DD2 & TMA & TM1 \\
  \hline
  APR4 & ---& 0.3 & 2.0 & 2.7 & 3.1 \\
  SFHo & 0.3 & ---& 1.7 & 2.5 & 3.0 \\
  DD2  & 2.0 & 1.7& --- & 1.2 & 2.3 \\
  TMA  & 2.7 & 2.5& 1.2 & --- & 1.5 \\
  TM1  & 3.1 & 3.0& 2.3 & 1.5 & --- 
 \end{tabular}
\\\vspace{5mm}
 \begin{tabular}{c|ccccc}
  0.03--2\,kHz& APR4 & SFHo & DD2 & TMA & TM1 \\
  \hline
  APR4 & ---  & 0.7  & 2.3 & 3.0 & 3.5 \\
  SFHo & 0.7  & ---  & 2.1 & 2.8 & 3.3 \\
  DD2  & 2.3  & 2.1  & --- & 1.6 & 2.5 \\
  TMA  & 3.0  & 2.8  & 1.6 & --- & 1.7 \\
  TM1  & 3.5  & 3.3  & 2.5 & 1.7 & --- 
 \end{tabular}
\hspace{1cm}
 \begin{tabular}{c|ccccc}
  0.05--2\,kHz& APR4 & SFHo & DD2 & TMA & TM1 \\
  \hline
  APR4 & ---  & 0.6  & 2.2 & 2.8 & 3.2 \\
  SFHo & 0.6  & ---  & 1.9 & 2.6 & 3.1 \\
  DD2  & 2.2  & 1.9  & --- & 1.5 & 2.4 \\
  TMA  & 2.8  & 2.6  & 1.5 & --- & 1.7 \\
  TM1  & 3.2  & 3.1  & 2.4 & 1.7 & --- 
 \end{tabular}
\\\vspace{5mm}
 \begin{tabular}{c|ccccc}
  0.03--4\,kHz& APR4 & SFHo & DD2 & TMA & TM1 \\
  \hline
  APR4 & ---  & 0.8  & 2.4 & 3.0 & 3.5 \\
  SFHo & 0.8  & ---  & 2.1 & 2.8 & 3.3 \\
  DD2  & 2.4  & 2.1  & --- & 1.7 & 2.6 \\
  TMA  & 3.0  & 2.8  & 1.7 & --- & 1.9 \\
  TM1  & 3.5  & 3.3  & 2.6 & 1.9 & --- 
 \end{tabular}
\hspace{1cm}
 \begin{tabular}{c|ccccc}
  0.05--4\,kHz& APR4 & SFHo & DD2 & TMA & TM1 \\
  \hline
  APR4 & ---  & 0.8  & 2.2 & 2.8 & 3.3 \\
  SFHo & 0.8  & ---  & 2.0 & 2.6 & 3.1 \\
  DD2  & 2.2  & 2.0  & --- & 1.6 & 2.4 \\
  TMA  & 2.8  & 2.6  & 1.6 & --- & 1.8 \\
  TM1  & 3.3  & 3.1  & 2.4 & 1.8 & --- 
 \end{tabular}
\end{table*}

\subsection{Analysis with the hybrid waveforms}

\begin{figure*}[t]
\begin{center}
\includegraphics[width=88mm]{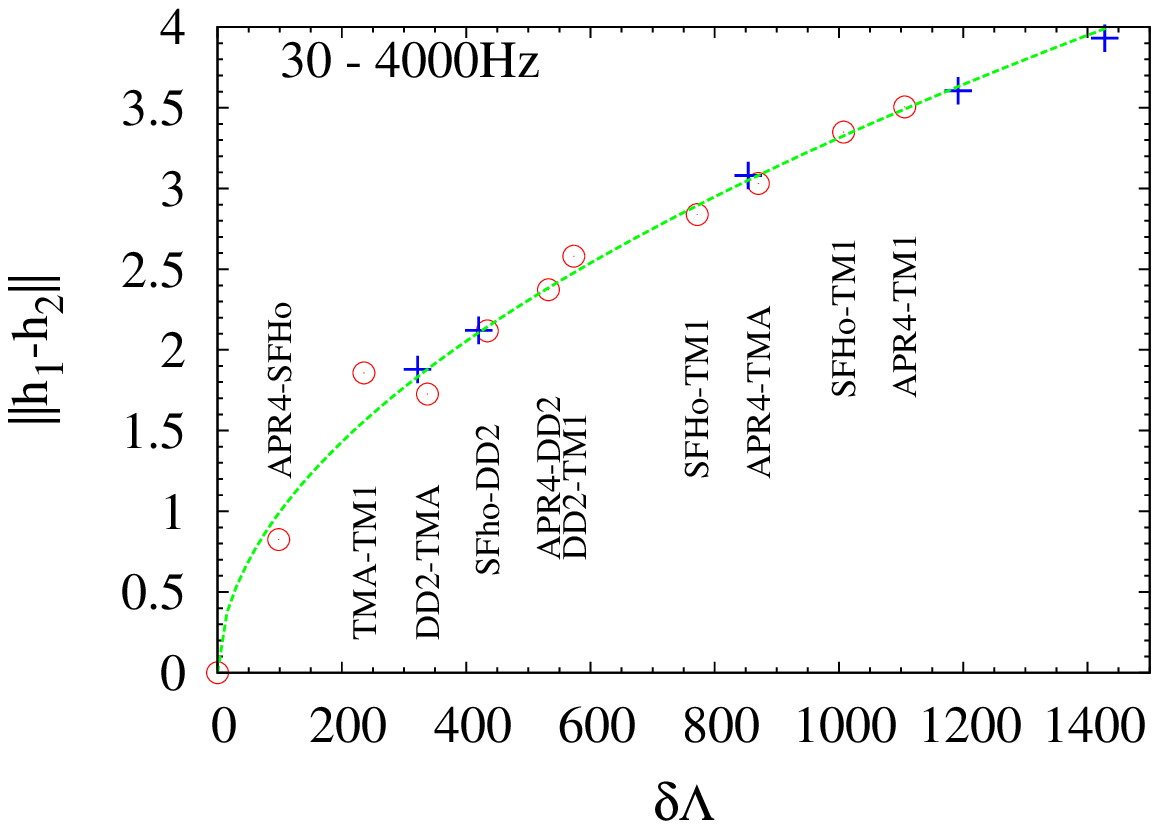}~~
\includegraphics[width=88mm]{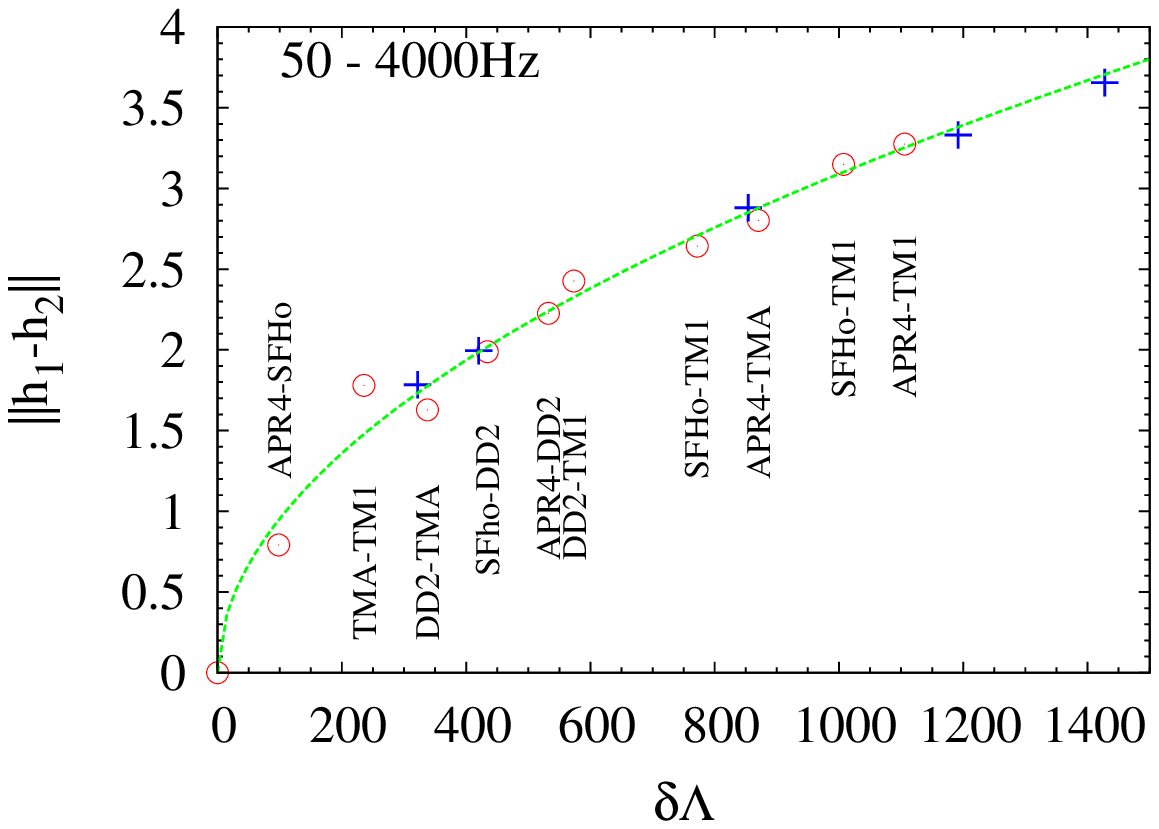}
\caption{$||h_1-h_2||$ for the hybrid waveforms as a function of
  $\delta\Lambda=|\Lambda_1-\Lambda_2|$ with $D_{\rm eff}=200$\,Mpc
  (open circles).  Left and right panels show for $(f_i,
  f_f)=$(30\,Hz, 4\,kHz) and (50\,Hz, 4\,kHz), respectively.  The
  values of $||h_1-h_2||$ are proportional to $200\,{\rm Mpc}/D_{\rm
    eff}$.  The dashed curve in each plot is a fitting formula in the
  form $||h_1-h_2||=A(\delta\Lambda/1000)^b$ where $(A, b)=(3.31,
  0.522)$ and $(3.09, 0.511)$ for the left and right panels,
  respectively. The labels like ``APR4-SFHo'' show the combination of
  two equations of state for each value of $\delta \Lambda$. The
  crosses denote the results of $||h_1-h_2||$ for the combination of
  binary neutron stars and binary black holes of mass
  $1.35$--$1.35M_\odot$. 
\label{fig4}}
\end{center}
\end{figure*}

Table~\ref{table2} lists the values of $||h_1-h_2||$ for all the
combination of the hybrid waveforms with the five equations of state
for an event of $D_{\rm eff}=200$\,Mpc.  Figure~\ref{fig4} also plots
$||h_1-h_2||$ as a function of $\delta\Lambda=|\Lambda_1-\Lambda_2|$
for $f_i=30$\,Hz or 50\,Hz and $f_f=4$\,kHz. Here, $\delta \Lambda$
denotes the absolute value in the difference of the dimensionless
tidal deformability of two different equations of
state. Table~\ref{table2} shows that the values of $||h_1-h_2||$
depend very weakly on the value of $f_f$ as long as it is larger than
2\,kHz. Furthermore, for $f_f=2$ and 4\,kHz, the values of
$||h_1-h_2||$ are only slightly [by (0.1--0.4)$\times (D_{\rm
    eff}/200\,{\rm Mpc})^{-1}$] larger than those for $f_f=1$\,kHz. It
is also found that the difference is large for the combination of two
soft equations of state.  All these results agree totally with the
results in Appendix~C, and hence, we may conclude that they hold
universally irrespective of the model waveforms.  From these results,
we confirm that the measurability is determined primarily by the late
inspiral waveform, and the contribution of the merger and post-merger
waveforms is minor.

Table~\ref{table2} also shows that for given combination of two
waveforms, the values for $f_i=50$\,Hz are by $\sim 10$\% smaller than
those for $f_i=30$\,Hz.  This also agrees quantitatively with the
results in the analysis in terms of Taylor-F2 approximant (see
Appendix~C), and hence, we could suppose that the values of
$||h_1-h_2||$ for $f_i=30$\,Hz would be only by $\sim 5$\% smaller
than those for $f_i=10$\,Hz. Nevertheless, they depend slightly on
the value of $f_i$.  This implies that the measurability of the tidal
effect is determined not only by the late inspiral waveform but also
by the relatively early one. 

We also calculated $||h_1-h_2||$ using another hybrid waveforms
derived with $(t_i, t_f)=$(10\,ms, 25\,ms): see Eq.~(\ref{corre}).  We
confirmed that the results depend only weakly on the choice of $t_i$
and $t_f$: Specifically, the change in the values of $||h_1-h_2||$
shown in Table~\ref{table2} is within 0.1 irrespective of the
waveforms. 

Figure~\ref{fig4} shows that for $\delta \Lambda \agt 100$, 400, and
800, $||h_1-h_2||$ is larger than 1, 2, and 3, respectively, for
$D_{\rm eff}=200$\,Mpc (note that if $f_i=10$\,Hz, the values of
$||h_1-h_2||$ would be by $\sim 5$\% and 15\% larger than those for
$f_i=30$\,Hz and 50\,Hz, respectively: see Appendix~C).  This implies
that for an event of $D_{\rm eff}=200$\,Mpc, two equations of state
are marginally distinguishable by the observation of inspiral and
merger waveforms by advanced gravitational-wave detectors if $\delta
\Lambda \agt 100$.

The neutron-star radius approximately monotonically increases with
$\Lambda$.  For the five equations of state employed in this paper,
the radius of $1.35M_\odot$ neutron stars is written as
\beqn
R_{1.35}=(13.565 \pm 0.076)\,{\rm km}
\left({\Lambda \over 1000}\right)^{0.16735 \pm 0.0094},\nonumber \\
\eeqn
where the standard errors shown for this fitting formula are at 
1-$\sigma$ level. By taking the variation, the relative difference in
the radius, $\delta R_{1.35}$, is related to $\delta \Lambda$ by
\beqn
\delta R_{1.35}=(0.91 \pm 0.05)\,{\rm km}
\left({R_{1.35} \over 13\,{\rm km}}\right)
\left({\delta\Lambda \over 400}\right)
\left({\Lambda \over 1000}\right)^{-1}. \nonumber \\
\eeqn
For stiff equations of state that yield a large neutron-star radius of
$R_{1.35} \agt 13.5$\,km, i.e., $\Lambda \agt 1000$, $\delta \Lambda$
for two different equations of state can become larger than $\sim 400$
if the difference in $R_{1.35}$ is larger than $\approx 0.9$\,km.
Thus, if the true equation of state is stiff, the equation of state
will be strongly constrained for an event of the advanced detectors at
$D_{\rm eff} \alt 200$\,Mpc, by which the measurability of $\delta
\Lambda$ is $\approx 400$ at 2-$\sigma$ level. 

By contrast, among soft equations of state, the difference in
$\Lambda$ is not as large as 400 (for the typical neutron-star mass
1.30--1.40$M_\odot$). For example, $\delta\Lambda$ for APR4 and SFHo
equations of state is $\sim 100$ although the difference in radius is
$\approx 0.8$\,km for neutron stars of mass $1.35M_\odot$ in these
equations of state. This implies that it will not be easy to
accurately identify the true equation of state among many candidate
soft equations of state for a typical advanced-LIGO event at $D_{\rm
  eff}=200$\,Mpc.  The reason for this is quite simple: The phase
difference between two waveforms for two different soft equations of
state can be appreciable only for a high-frequency range of $\agt
1$\,kHz, for which the sensitivity of the operating and planned
gravitational-wave detectors is not very high (see
Fig.~\ref{fig3}). This situation cannot be significantly improved even
if we take into account the merger and post-merger waveform, because
gravitational waves in these phases have a high frequency, and do not
contribute a lot to enhancing the signal-to-noise ratio, as shown in
Table~\ref{table2}.  However, even if the true equation of state is
soft, it will be still possible to discriminate it from stiff
equations of state that yield $\Lambda \agt 1000$. Thus, the detection
of gravitational waves emitted at $D_{\rm eff} \alt 200$\,Mpc for the
advanced detectors will give us an impact even if the true equation of
state is soft. We also should mention that if we fortunately have a
nearby event at $D_{\rm eff} \ll 200$\,Mpc, the situation will become
much more optimistic.

\begin{table}[t]
 \caption{The same as Table~\ref{table2} but in between the 
   hybrid waveforms for binary neutron stars and binary black holes.
\label{table2a}}
 \begin{tabular}{l|ccccc}
    & APR4 & SFHo & DD2 & TMA & TM1 \\
  \hline
~~  0.03--1\,kHz~~ & 1.4 & 1.8 & 2.9 & 3.4 & 3.8 \\
~~  0.03--2\,kHz~~ & 1.9 & 2.1 & 3.1 & 3.6 & 3.9 \\
~~  0.03--4\,kHz~~ & 1.9 & 2.1 & 3.1 & 3.6 & 3.9 \\
~~  0.05--1\,kHz~~ & 1.3 & 1.6 & 2.7 & 3.2 & 3.5 \\
~~  0.05--2\,kHz~~ & 1.8 & 1.9 & 2.8 & 3.3 & 3.6 \\
~~  0.05--4\,kHz~~ & 1.8 & 2.0 & 2.9 & 3.3 & 3.7 \\
 \end{tabular}
\end{table}

Next, we evaluate $||h_1-h_2||$ employing hybrid waveforms for binary
neutron stars and spinless binary black holes of mass
$1.35$--$1.35M_\odot$ assuming $D_{\rm eff}=200$\,Mpc. For this
analysis, a hybrid waveform for the binary black hole is constructed
by combining a numerical waveform and an EOB one as we already did for
binary neutron stars. Here, the numerical waveform is again taken from
SXS Gravitational Waveform Database~\cite{SXS} and we employ
SXS:BBH:001.  Table~\ref{table2a} lists the results of $||h_1-h_2||$
and in Fig.~\ref{fig4}, we plot the data setting $\Lambda=0$ for the
black-hole case (see the crosses). These show that $||h_1-h_2|| \agt
2$ for $D_{\rm eff}=200$\,Mpc irrespective of the neutron-star
equations of state we employ.  This indicates that gravitational waves
from binary neutron stars for $D_{\rm eff} \alt 200$\,Mpc will be
distinguished from those from binary black holes of the same mass with
a certain confidence level.  

Figure~\ref{fig4} also shows that the relation between $||h_1-h_2||$
and $\delta \Lambda$, satisfied for binary neutron stars, is
approximately satisfied even for the waveform combination of binary
neutron stars and binary black holes.  This also indicates that
gravitational waves from binary neutron stars for $D_{\rm eff} \alt
200$\,Mpc will be distinguished from those from binary black holes at
2-$\sigma$ level if the value of $\Lambda$ for the neutron stars is
larger than $\sim 400$.

Before closing this subsection, we note the following point.  By
comparing our results with those in Ref.~\cite{read13}, it is found
that our results for the measurability of $\delta\Lambda$ and
$R_{1.35}$ agree approximately with theirs. However, this is
accidental. In Ref.~\cite{read13}, the measurability was explored
choosing $f_i$ in Eq.~(\ref{dSNR}) to be 200\,Hz while we choose it to
be 30\,Hz. 
As found from Table~\ref{table2} (see also Appendix~C), the values of
$||h_1-h_2||$ systematically decrease with the increase of $f_i$ for a
given value of $f_f$. We checked that for $f_i=200$\,Hz, the values of
$||h_1-h_2||$ could be half of those for $f_i=30$\,Hz.  This implies
that our results, based on new hybrid waveforms, actually show weaker
measurability than that in Ref.~\cite{read13}.  The precise reason is
not very clear. However, it is reasonable to speculate that in the
previous work, the numerical dissipation and the absence of any
appropriate extrapolation procedure result in spuriously shorter
merger time even for the highest-resolution runs as shown
in~Refs.~\cite{Hotokezaka2013,hotoke2015}, so that the tidal effects
could be spuriously overestimated. In addition, as noted in
Ref.~\cite{read13}, the systematic error in their hybrid waveforms
might be non-negligible because of a small number of the wave cycles
and large initial residual eccentricity: These errors would also
systematically enhance the measurability of the tidal deformability of
Ref.~\cite{read13}. 

\subsection{Validity of analytic/semi-analytic waveforms}

We then evaluate $||h_1-h_2||$ choosing the hybrid waveforms as $h_1$
while the EOB, hybrid-TT4, and TF2 waveforms as $h_2$,
respectively. 
Here, as the EOB waveforms, we only take into account the inspiral
part. Note that in the EOB formalism we employ in this paper, the
amplitude approaches zero if the orbital separation approaches zero.
The ``hybrid-TT4'' waveforms are constructed by combining the
numerical and TT4 waveforms, and then the Fourier transformation is
performed straightforwardly. The TF2 approximant that we employ in
this paper is described in Appendix~C.

The purpose of this analysis is to assess how appropriate the
EOB/hybrid-TT4/TF2 waveforms are as {\em inspiral} model gravitational
waveforms.  We note that for the EOB and TF2 waveforms employed, the
spectrum with $f \agt 1$\,kHz is not very realistic because of the
absence of the merger and post-merger waveforms, and hence, it is not
appropriate to take the higher-frequency part into consideration for
the comparison with the hybrid waveforms.  Also, as we already showed
in Sec.~IV A, the values of $||h_1-h_2||$ for $(f_i, f_f)=(50\,{\rm
  Hz}, 1\,{\rm kHz})$ are only by $\sim 0.2$ smaller than those for
$(f_i, f_f)=(30\,{\rm Hz}, 1\,{\rm kHz})$ for an event of advanced
LIGO at $D_{\rm eff}=200$\,Mpc.  Thus, in this section, all the
analyses will be performed choosing $(f_i, f_f)=(50\,{\rm Hz}, 1\,{\rm
  kHz})$ for simplicity.

Three panels of Table~\ref{table3} list the values of
$||h_1-h_2||$ for $f_i=50$\,Hz and $f_f=1$\,kHz for the combination of
the hybrid and other waveforms assuming $D_{\rm eff}=200$\,Mpc. 
From the comparison between Tables~\ref{table2} and~\ref{table3}, it
is found that for APR4, SFHo, DD2, and TMA equations of state, the EOB
waveforms can reproduce approximately the same results of
$||h_1-h_2||$ (within the error of $\pm 0.2$) as for the hybrid
(hybrid-EOB) waveforms. This fact makes us confirm again that the EOB
formalism would have already become robust for generating accurate
inspiral waveform templates if the neutron-star equation of state is
not very stiff, i.e., $\Lambda$ is smaller than $\sim 1000$.
By contrast, the EOB waveforms may not be yet accurate enough for
neutron stars with very stiff equations of state. For TM1, this fact
is in particular noticeable: It is clearly found from the result of
$||h_1-h_2||$ for the combination of the TM1 EOB and TM1 hybrid
waveforms, which is significantly different from zero. This suggests
again that for very stiff equations of state, there is still a room
for improving the EOB formalism~\cite{tania}.

\begin{table}[t]
 \caption{The same as Table~\ref{table2} but between the hybrid and
   EOB waveforms (upper table), between the hybrid and hybrid-TT4
   waveforms (middle table), and between the hybrid and TF2 waveforms
   (bottom table). $f_i=50$\,Hz and $f_f=1$\,kHz are chosen. 
\label{table3}}
\hspace{7.9mm}
 \begin{tabular}{l|ccccc}
  0.05--1\,kHz& APR4 & SFHo & DD2 & TMA & TM1 \\
  \hline
  EOB:APR4 & 0.2& 0.3& 1.9 & 2.6 & 3.1 \\
  EOB:SFHo & 0.5& 0.2& 1.6 & 2.4 & 2.9 \\
  EOB:DD2  & 2.0& 1.7& 0.2 & 1.3 & 2.2 \\
  EOB:TMA  & 2.8& 2.6& 1.2 & 0.3 & 1.5 \\
  EOB:TM1  & 3.1& 3.0& 2.0 & 0.9 & 0.9
 \end{tabular}
\\\vspace{0.5cm}
 \begin{tabular}{l|ccccc}
  \hspace{10mm}0.05--1\,kHz& APR4 & SFHo & DD2 & TMA & TM1 \\
  \hline
 hybrid- TT4:APR4 & 0.2 & 0.5 & 2.1 & 2.7 & 3.2 \\
 hybrid- TT4:SFHo & 0.2 & 0.2 & 1.9 & 2.6 & 3.0 \\
 hybrid- TT4:DD2  & 1.7 & 1.4 & 0.4 & 1.5 & 2.5 \\
 hybrid- TT4:TMA  & 2.3 & 2.1 & 0.7 & 0.6 & 1.9 \\
 hybrid- TT4:TM1  & 2.8 & 2.6 & 1.8 & 1.0 & 0.7
 \end{tabular}
\\\vspace{0.5cm}
\hspace{9.4mm}
 \begin{tabular}{l|ccccc}
  0.05--1\,kHz& APR4 & SFHo & DD2 & TMA & TM1 \\
  \hline
  TF2:APR4 & 0.3& 0.4& 2.1 & 2.7 & 3.2 \\
  TF2:SFHo & 0.4& 0.3& 1.8 & 2.5 & 3.1 \\
  TF2:DD2  & 1.9& 1.6& 0.4 & 1.5 & 2.4 \\
  TF2:TMA  & 2.7& 2.5& 1.0 & 0.5 & 1.7 \\
  TF2:TM1  & 3.0& 2.9& 1.8 & 0.6 & 1.1
 \end{tabular}
\end{table}

\begin{figure*}[t]
\begin{center}
\includegraphics[width=84mm]{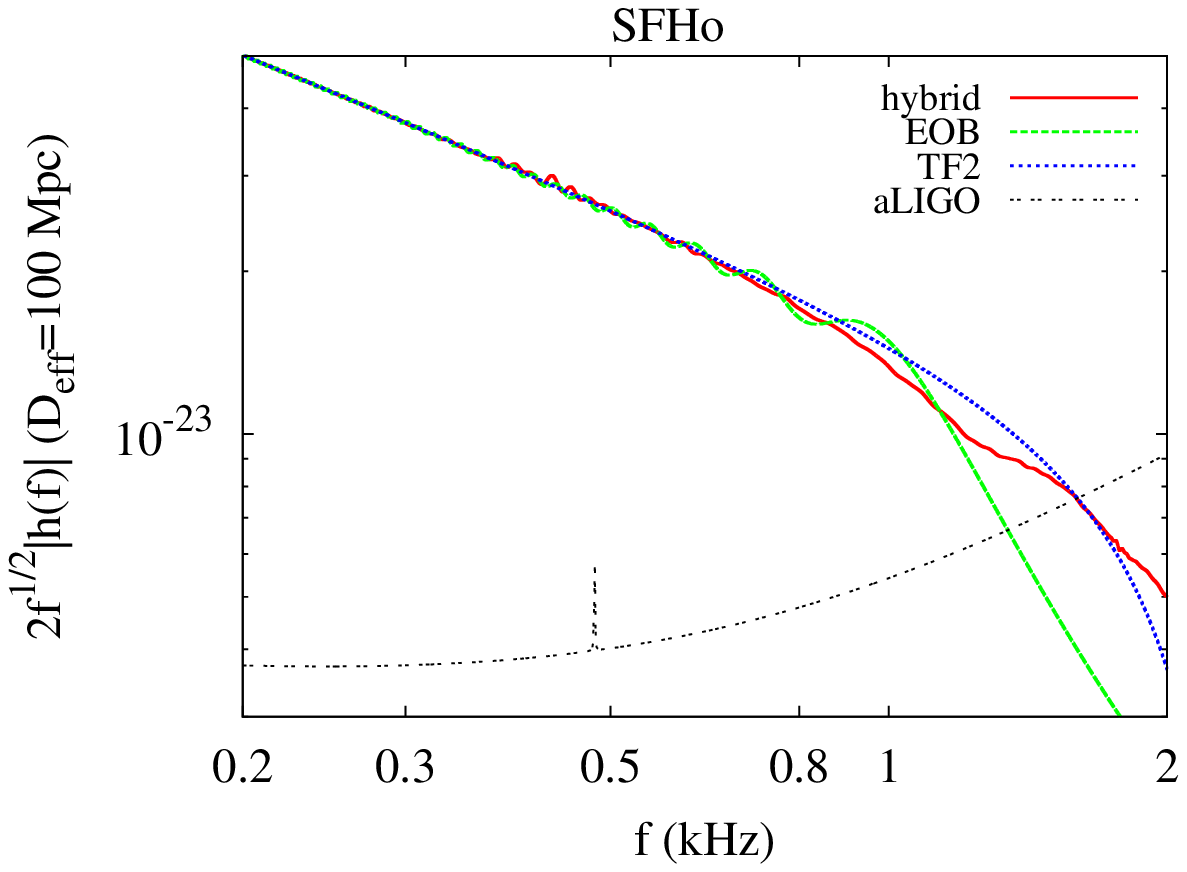}~~~
\includegraphics[width=84mm]{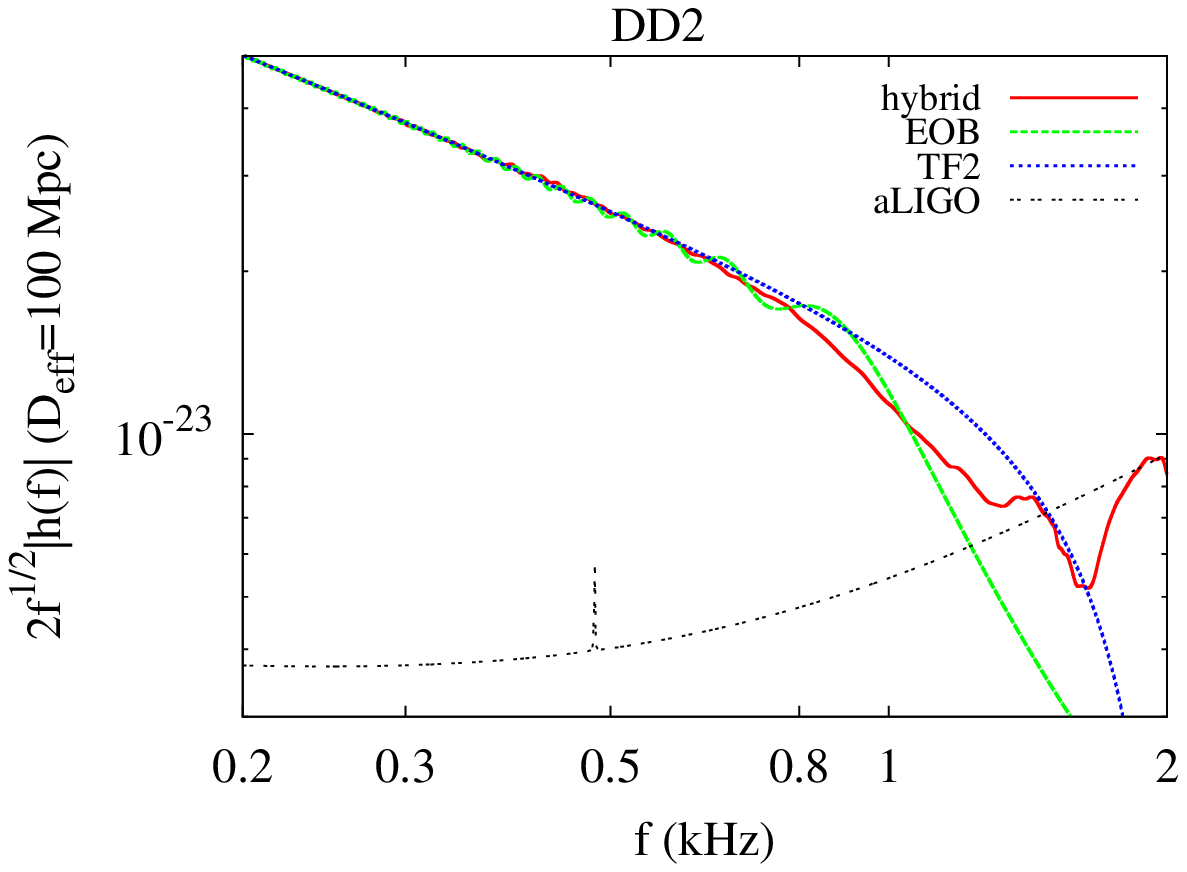}\\
\includegraphics[width=84mm]{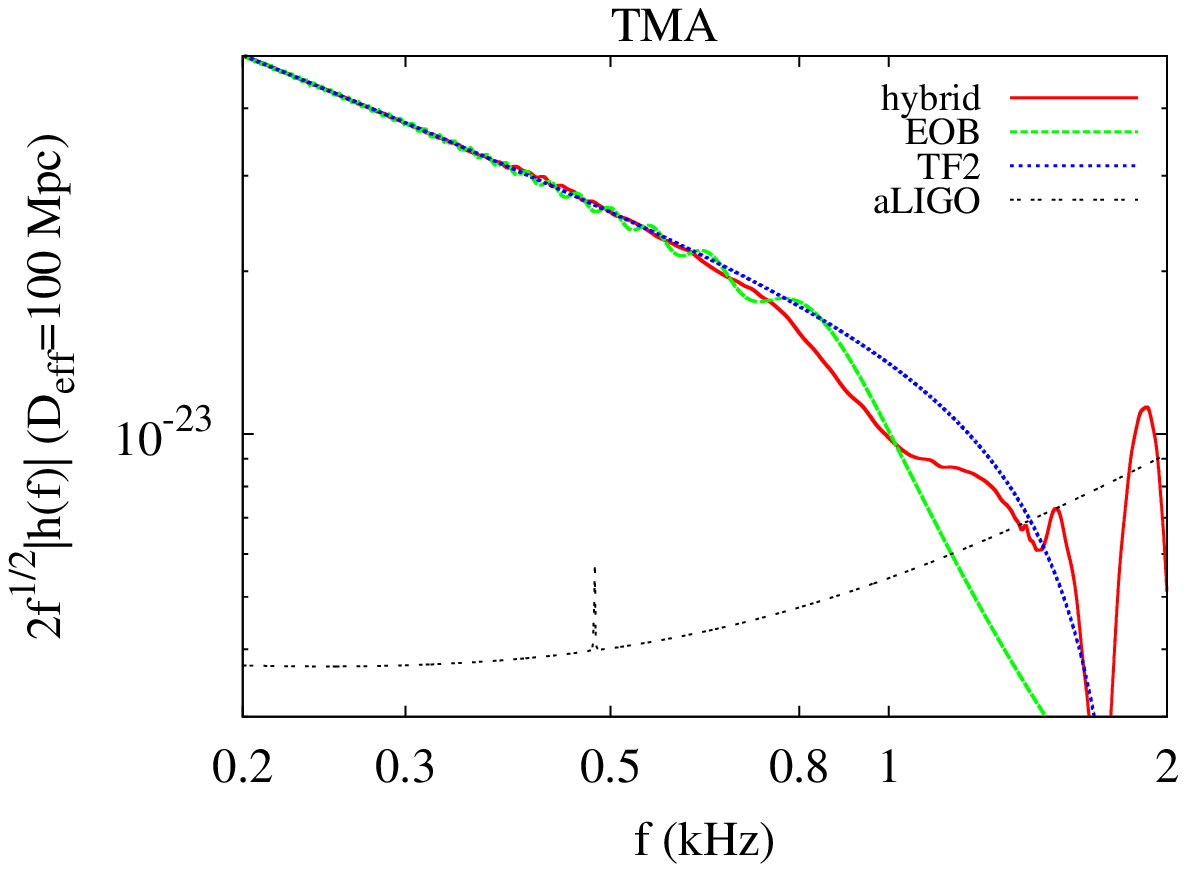}~~~
\includegraphics[width=84mm]{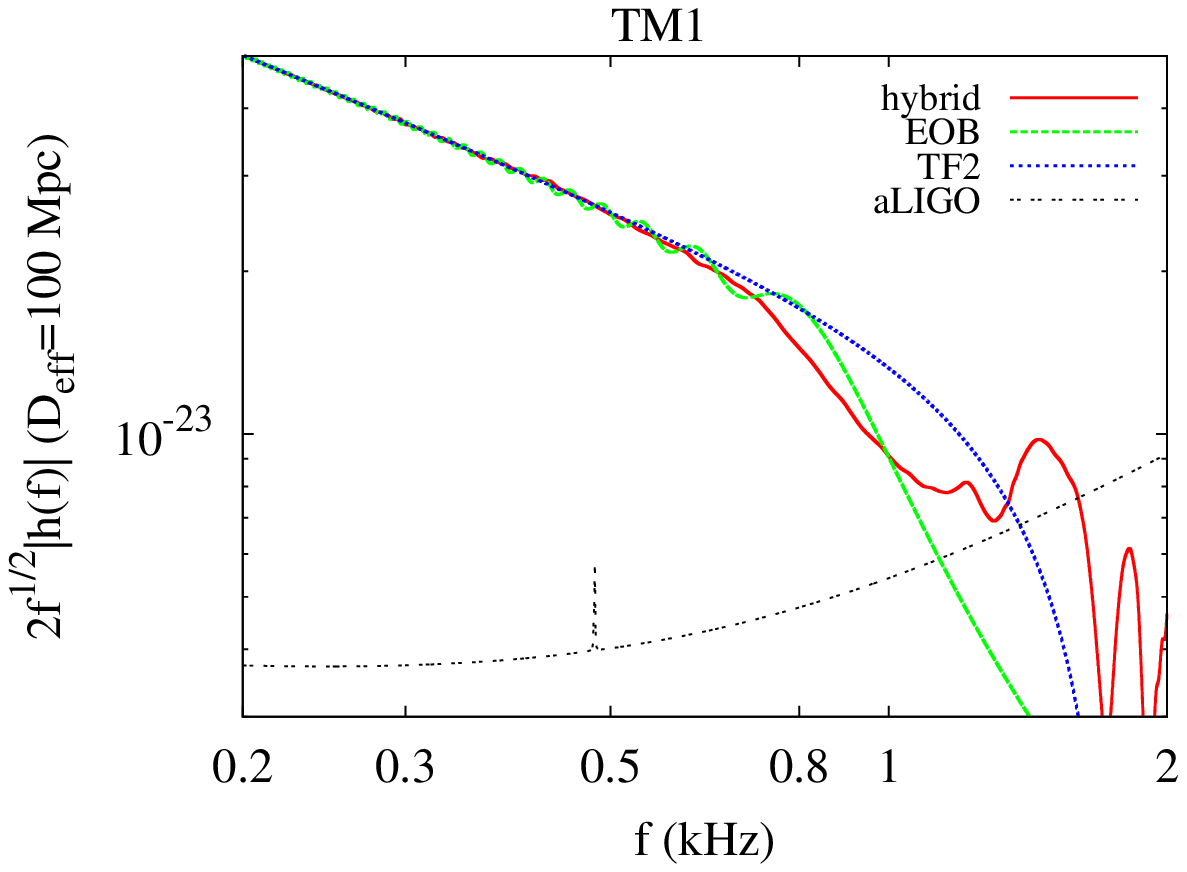}
\caption{Comparison of the spectrum shapes of hybrid, Taylor-F2 (TF2),
  and EOB waveforms for SFHo (upper left), DD2 (upper right), TMA
  (lower left) and TM1 (lower right) equations of state. 
\label{fig5}}
\end{center}
\end{figure*}

\begin{figure}[t]
\begin{center}
\includegraphics[width=86mm]{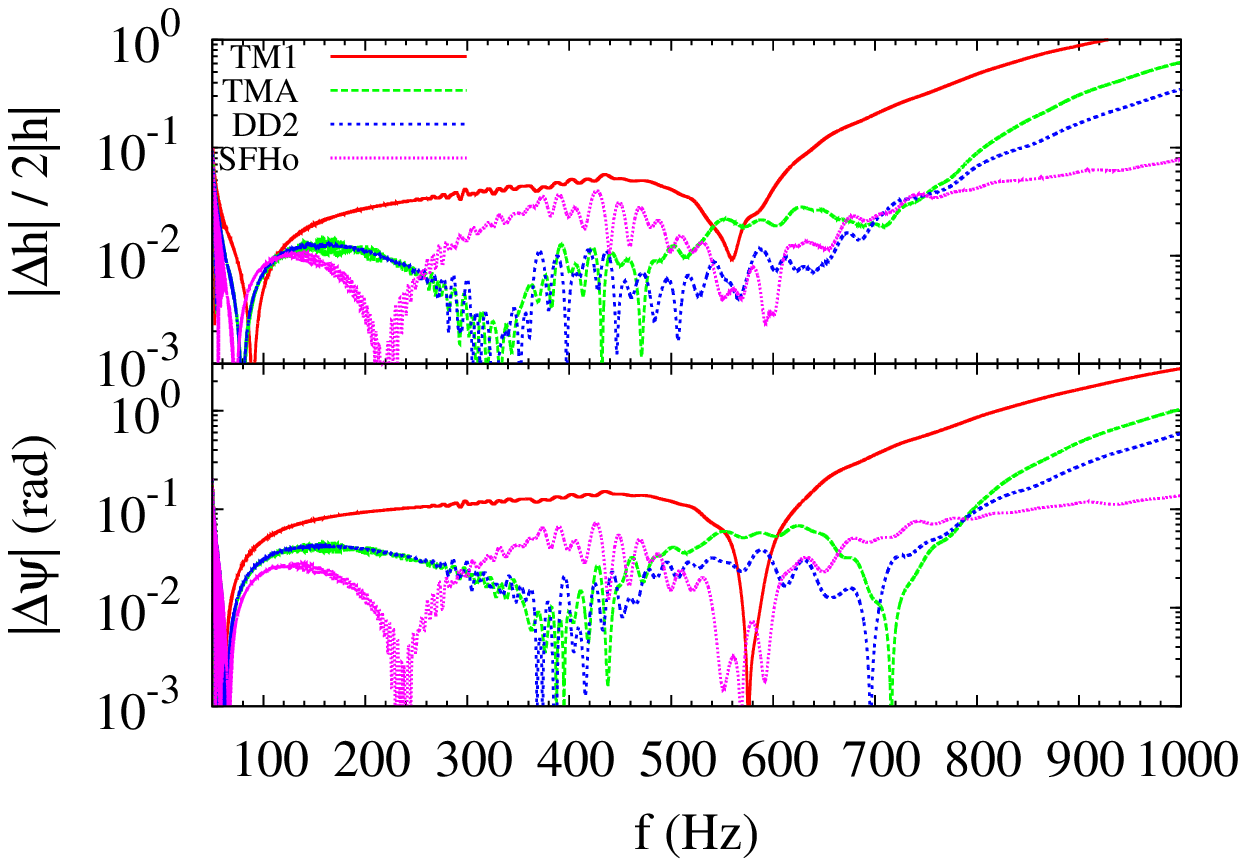}
\caption{Absolute values for the difference of the Fourier waveforms
  (upper panel) and Fourier phases (lower panel) as functions of the
  gravitational-wave frequency between the hybrid and TF2
  waveforms. Note that at the minima of $\Delta \psi$, its sign
  changes. 
\label{fig6}} 
\end{center}
\end{figure}

We also find from Table~\ref{table3} that the values of $||h_1-h_2||$
for the choice of the hybrid-TT4 or TF2 waveforms are more appreciably
different from those in Table~\ref{table2} (except for the APR4 and
SFHo equations of state; hybrid-EOB and hybrid-TT4 waveforms agree
with each other in a good manner for these equations of state).  This
fact is also found from, e.g., (i) the diagonal components in
Table~\ref{table3} (i.e., for the case that $h_1$ and $h_2$ for the
same equation of state are employed) is significantly different from
zero, in particular for stiff equations of state, (ii) asymmetry
between the off-diagonal components, which should be absent for the
templates, is more appreciable, and (iii) the hybrid-TT4 waveform for
the TM1 equation of state matches better with the hybrid (hybrid-EOB)
waveform for the TMA than for the TM1.  If the hybrid waveforms would
be more realistic ones, these results imply that hybrid-TT4 and TF2
waveforms would not be as good measurement templates as the EOB ones.
This also indicates that the templates by the TF2 and TT4 approximants
would give a systematic bias in the estimation of tidal deformability.
This agrees qualitatively with the finding in Ref.~\cite{wade14}.


One of the reasons for the disagreement between two hybrid waveforms
(hybrid-EOB and hybrid-TT4) is that the effect of the tidal
deformation would be underestimated in the current TT4 approximant,
due to the lack of higher-order PN terms (see Sec.~III and
Appendix~B). Another reason is that the matching frequency in our
present study ($f \sim 400$\,Hz) would be still high: For such
frequency, the EOB and TT4 waveforms do not agree well with each other
for high values of $\Lambda$ and the phase difference is not
negligible; for stiff equations of state, the accumulated phase
difference is $\sim 0.3(\Lambda/1000)$\,rad for 50\,Hz $\leq f \leq
400$\,Hz (see Appendix~B).  The phase difference that results from the
incompleteness of the tidal effects would be proportional
approximately to $\Lambda f_f^{\alpha}$ where $\alpha\geq 8/3$ [see
  Appendix~B and Eqs.~(\ref{eqTF2}) and (\ref{eqAA})] and $f_f$ is the
upper end of the matching frequency. Thus, if the hybridization is
performed with a lower value of the matching frequency, the
disagreement between two hybrid waveforms would be smaller.  Our
numerical results (compare Figs.~\ref{fig1} and \ref{fig2}) support
that the EOB waveforms would be more accurate to perform matching at
$f \sim 400$\,Hz than the TT4 ones.  However, to confirm these
speculations, we will have to perform a longer-term simulation and
have to match the waveform at lower frequency.

One reason that the current version of the TF2 approximant does not
reproduce the hybrid waveforms is found from the analysis of the
spectrum shape.  Figure~\ref{fig5} plots the Fourier spectra of the
three different models (hybrid, pure EOB, and TF2) for SFHo, DD2, TMA,
and TM1 equations of state.  In the spectrum of the EOB waveform, a
modulation is found.  This is due to the fact that the time-domain
waveform is artificially terminated at the end of the inspiral phase
and hence the spectrum is subject to the Gibbs phenomenon.  Besides
this modulation, the spectrum shapes of the hybrid and EOB waveforms
are in a fair agreement for $f \alt 1$\,kHz. This should be the case
because the agreement between the two waveforms has been already found
in particular for the equations of state with $\Lambda < 1000$ (see
Fig.~\ref{fig1}).  By contrast, the spectrum amplitude of the TF2
approximant does not agree well with those of the hybrid waveforms for
the late inspiral phase ($f \agt 500$\,Hz) in which the tidal effects
as well as general relativistic gravity play an important role: The
steep decline of the spectrum observed in the hybrid waveforms for $f
\agt 500$\,Hz cannot be well captured by the current version of the
TF2 approximant in particular for the stiff equations of state like
TMA and TM1. This indicates that the tidal effects would not be
sufficiently taken into account in this TF2 spectrum amplitude.  (We
note that this insufficiency is partly due to the use of the
stationary phase approximation.)

The phases of the Fourier transform in the hybrid and TF2 waveforms also
do not agree well with each other. Figure~\ref{fig6} plots the
absolute difference (upper panel) and phase difference (lower panel)
between the hybrid and TF2 waveforms for given equations of
state. Here the absolute difference of the waveforms for a given value
of $f$ is defined by
\beqn
{|\tilde h_1(f)-\tilde h_2(f)| \over 2|\tilde h_1(f)|},
\eeqn
where $\tilde h_1$ and $\tilde h_2$ denote the Fourier transform of
the hybrid and TF2 waveforms.  For plotting Fig.~\ref{fig6}, we choose
$\Delta t$ and $\Delta \phi$ that minimize $||h_1-h_2||$ of
Eq.~(\ref{dSNR}) for $f_i=50$\,Hz and $f_f=1$\,kHz.  This figure shows
that the absolute difference in the waveform is determined primarily
by the phase difference and that the phase difference is generally
larger for larger values of $\Lambda$.  This suggests that the absence
of higher-order PN terms in the tidal-deformation effect would be one
of the primary sources for the disagreement in the phase.


We also note that the phase difference is present rather uniformly for
50--1000\,Hz even for soft equations of state like SFHo for which the
tidal-deformation effect should be minor. This suggests that the
absence of not only the tidal effect but also other non-tidal
higher-order PN terms like 4PN and higher-order terms would cause
inaccuracy of the TF2 approximants. A recent study for the extension
of the TF2 approximant in the context of binary black
holes~\cite{khan15} indeed suggests that the coefficients of the absent
higher-order PN terms in phase ($\psi_{\rm TF2}$: see Appendix~C)
should be large (the order of $\alpha_k$ with $k \geq 9$ in
Eq.~(\ref{eqTF2}) would of $10^4$ or more, i.e., comparable to the
tidal-effect terms) perhaps due to the use of the stationary phase
approximation, and this should affect the wave phase in the late
inspiral stage.  Therefore, for improving the performance of the TF2
approximant, we will have to incorporate both the tidal and non-tidal
higher-order PN terms, which are absent in the current version. We
plan to explore this issue in the future work.

\section{Summary}

Combining new gravitational waveforms derived by long-term (14--16
orbits) numerical-relativity simulations with the waveforms by an EOB
formalism for coalescing binary neutron stars, we constructed hybrid
waveforms and estimated the measurability for the dimensionless tidal
deformability of the neutron stars, $\Lambda$, by ground-based
advanced gravitational-wave detectors, using the hybrid waveforms as
the model waveforms.  We found that for an event at a hypothetical
effective distance of $D_{\rm eff}=200$\,Mpc, the distinguishable
difference in the dimensionless tidal deformability for $1.35M_\odot$
neutron stars will be $\approx 100$, 400, and 800 at 1-$\sigma$,
2-$\sigma$, and 3-$\sigma$ levels, respectively, for the advanced
LIGO. If the true equation of state is stiff and the corresponding
neutron-star radius is $R \agt 13$\,km, this suggests that $R$ will
be constrained within $\approx 1$\,km at the 2-$\sigma$ level for an
event of $D_{\rm eff}=200$\,Mpc. On the other hand, if the true
equation of state is soft and $R \alt 12$\,km, it will be difficult to
accurately identify the equations of state among many soft candidates,
although it is still possible to discriminate it from stiff equations
of state with $R \agt 13$\,km. These results indicate that measuring
the tidal deformability is a promising method for constraining the
neutron-star equation of state in the near future.

The analysis in this paper was carried out for given values of mass
and mass ratio of the binaries. In reality, these parameters have to
be also determined in the data analysis.  The uncertainty in these
parameters will enhance the uncertainty in the estimation for the
dimensionless tidal deformability as shown in Ref.~\cite{wade14}.
Therefore, the estimation for the measurability of the dimensionless
tidal deformability in this paper would be optimistic. We are now
deriving more numerical waveforms changing the mass and mass
ratio. More realistic analysis for the measurability will be carried
out in the next work.

We also examined the validity of the waveforms by the EOB, TT4
(hybrid-TT4), and TF2 formalisms. Our analysis shows that these
waveforms deviate from our hybrid waveforms.  Comparison between the
hybrid waveforms and those by these approximants suggests that the EOB
waveform would be better than others.  However, there is still a room
for the improvement of the current EOB formalism in particular for neutron
stars with stiff equations of state in which $\Lambda > 1000$.  For
the current version of the TT4 and TF2 approximants, the absence of
higher-order PN terms is the likely source for the inaccuracy. For the
TT4, the absence of the higher-order PN terms in the tidal effects is
the main source for the inaccuracy. For the TF2, the absence of both
higher-PN terms in the tidal and non-tidal effects is likely to be the
source for the inaccuracy. Improving these approximants is one of the
interesting issues for the future.

\begin{acknowledgments}

We would like to thank the SXS collaboration for freely providing a
variety of high-precision gravitational waveforms of binary-black-hole
coalescence.  We thank H. Tagoshi and K. Kawaguchi for helpful
discussion.  This work was supported by a Grant-in-Aid for Scientific
Research (24244028), a Grant-in-Aid for Scientific Research on
Innovative Areas "New Developments in Astrophysics Through
Multi-Messenger Observations of Gravitational Wave Sources"
(15H00782), and a Grant-in-Aid for Research Activity Start-up 15H06857
of Japanese MEXT/JSPS.  KK was supported by the RIKEN iTHES project.

\end{acknowledgments}

\appendix

\begin{figure*}[t]
\begin{center}
\includegraphics[width=84mm]{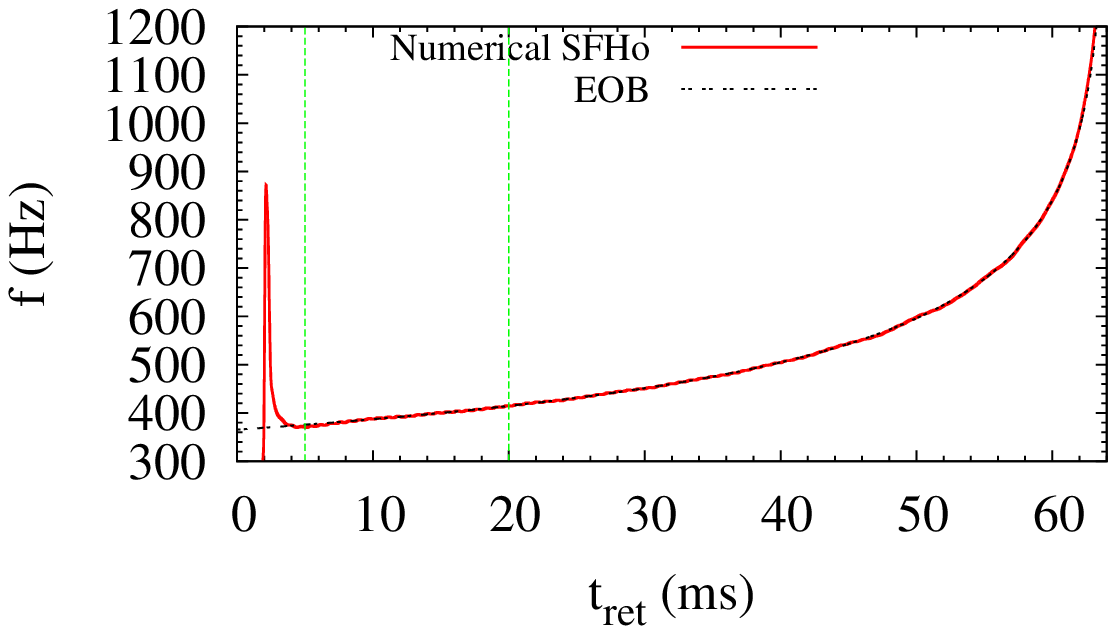}~~~~
\includegraphics[width=84mm]{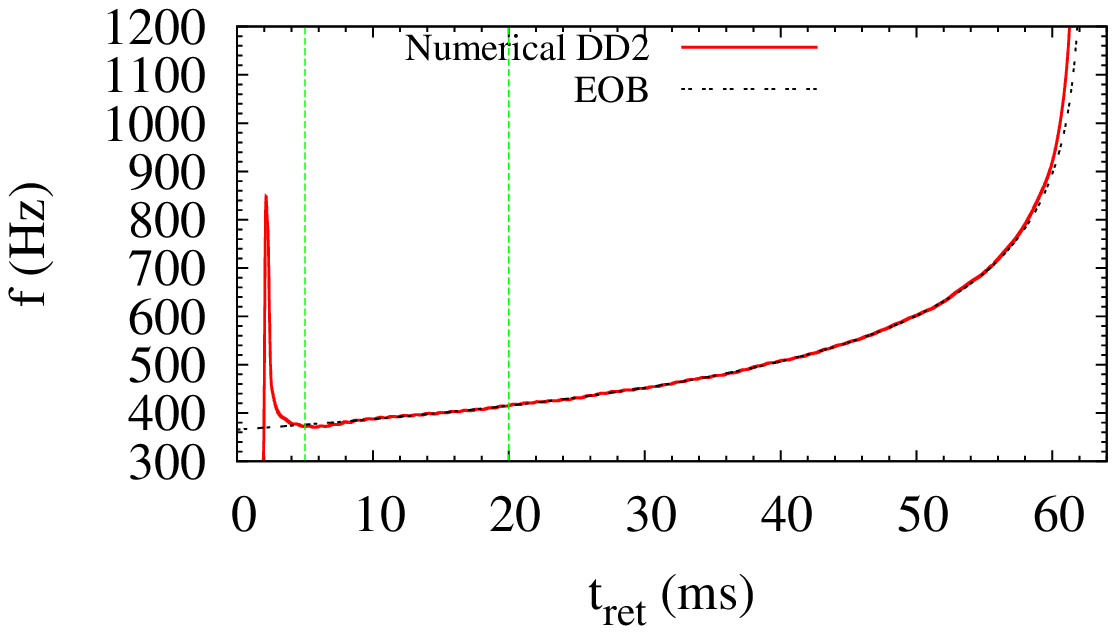}\\
\vspace{5mm}
\includegraphics[width=84mm]{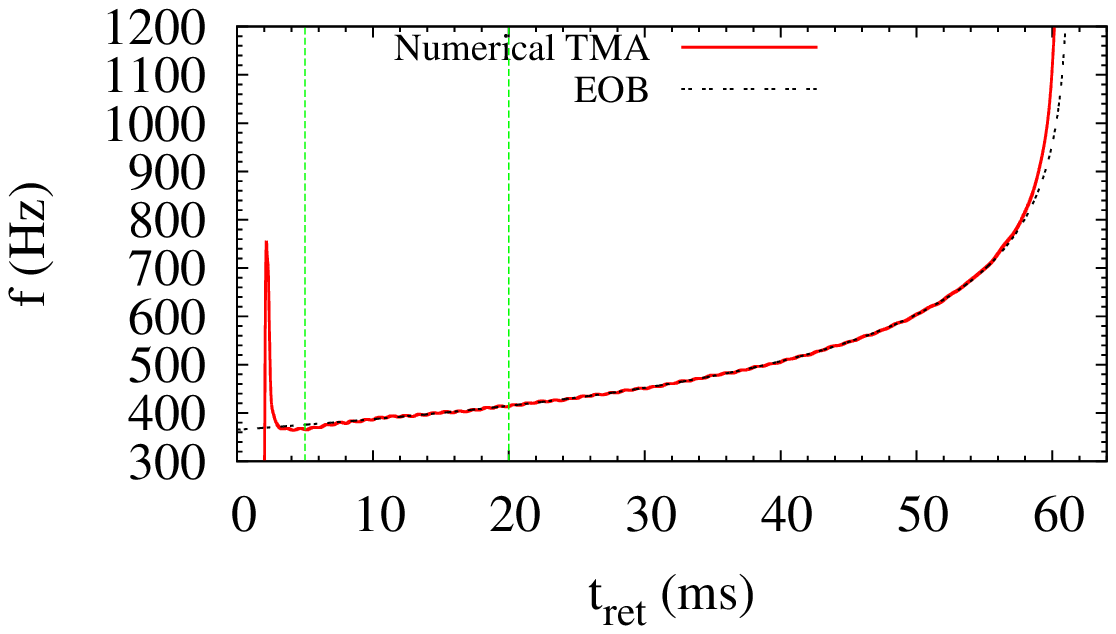}~~~~
\includegraphics[width=84mm]{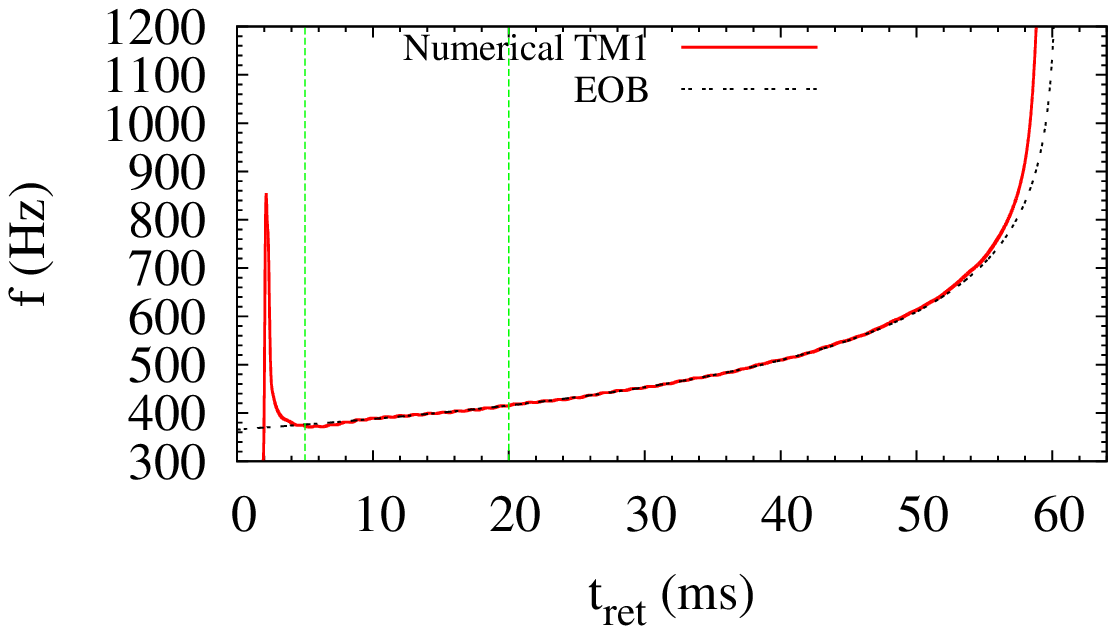}
\caption{Gravitational-wave frequency as a function of the retarded
  time for the SFHo (upper left panel), DD2 (upper right panel), TMA
  (lower left panel), and TM1 (lower right panel) equations of state.
  The solid and dot-dot curves denote the results of numerical and EOB
  waveforms, respectively. The vertical dashed lines show $t_{\rm
    ret}=5$\,ms and 20\,ms. The spike at $t_{\rm ret}\approx 2$\,ms is
  due to the unphysical modulation of the gravitational waveforms (see
  text).
\label{figap0}}
\end{center}
\end{figure*}

\section{Gravitational-wave frequency evolution}

For providing supplementary information of the gravitational waveforms
plotted in Fig.~\ref{fig1}, we show gravitational-wave frequency as a
function of the retarded time for numerical (solid curves) and EOB
(dot-dot curves) waveforms in Fig.~\ref{figap0}.  As we already
described in Ref.~\cite{hotoke2015}, two frequency curves agree with
each other except for $t_{\rm ret} \alt 5$\,ms and for the stage just
prior to the merger (around $t_{\rm ret} \sim 60$\,ms). The early-time
spike around $t_{\rm ret}\approx 2$\,ms and associated modulation are
caused by the fact that the initial condition, which describes
inspiral binary neutron stars only approximately because a conformal
flatness formulation is employed~(e.g.,~Ref.~\cite{TS2010}), is
contaminated by an unphysical component of gravitational waves. Thus,
the numerical waveforms only with $t_{\rm ret} \agt 5$\,ms are
reliable.  The late-time disagreement is larger for the stiff equation
of state which has high values of $\Lambda \agt 1000$, as expected
from Fig.~\ref{fig1}. This also indicates that there is still a room
for incorporating additional tidal effects into the EOB formalism for
improving it.  On the other hand, for softer equations of state with
$\Lambda < 1000$ like the SFHo equation of state, the disagreement is
minor. This indicates that the EOB waveforms well capture the
tidal-deformation effects as long as $\Lambda \ll 1000$.

\section{Comparison of the EOB and TT4 wave phases}

\begin{figure}[t]
\begin{center}
\includegraphics[width=88mm]{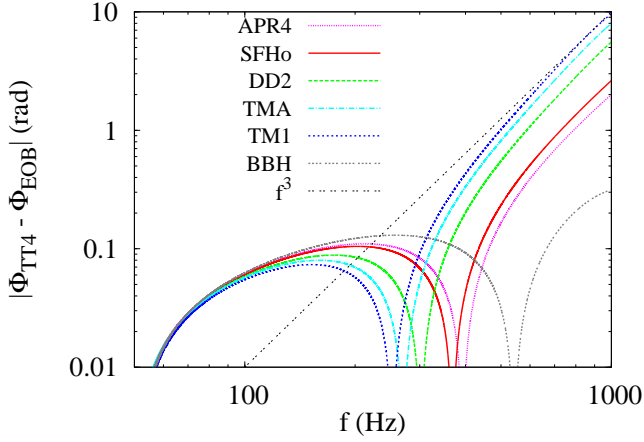}
\caption{Absolute value of the difference in the wave phases
  calculated by the TT4 approximant, $\Phi_{\rm TT4}$, and that of the
  EOB formalism, $\Phi_{\rm EOB}$, as a function of the
  gravitational-wave frequency, $f$. For $f \alt 200$\,Hz, $\Phi_{\rm
    EOB}$ is always larger than $\Phi_{\rm TT4}$ while for $f \agt
  400$\,Hz, $\Phi_{\rm TT4}$ is always larger than $\Phi_{\rm EOB}$
  for the binary-neutron-star models.  The dot-dot line shows
  $10(f/1\,{\rm kHz})^3$\,rad. The curve labeled by ``BBH'' is the
  case for $\Lambda=0$. 
\label{figap}}
\end{center}
\end{figure}

We compare the wave phases derived by an EOB and TT4 equation of
motion. Figure~\ref{figap} plots the absolute value of the difference
in the wave phases (wave phase of the TT4 approximant, $\Phi_{\rm
  TT4}$, minus that of the EOB formalism, $\Phi_{\rm EOB}$) as a
function of the gravitational-wave frequency, $f$, for the APR, SFHo,
DD2, TMA, and TM1 equations of state. For taking the difference, we
align the two phases at $f=50$\,Hz.  For $f \agt 400$\,Hz, $\Phi_{\rm
  TT4}$ is always larger than $\Phi_{\rm EOB}$, and the difference
steeply increases with $f$.  This is due to the fact that the orbital
(gravitational-wave frequency) evolution in the TT4 approximant is
slower than that in the EOB formalism in such a frequency band.  On
the other hand, for $f \alt 200$\,Hz, $\Phi_{\rm EOB}$ is by $\alt
0.1$\,rad larger than $\Phi_{\rm TT4}$. This would stem from the
difference in more than 4PN non-tidal terms between the EOB and the
TT4 equations of motion. To clarify this fact, we also plot the curve
for $\Lambda=0$ (see the curve labeled by ``BBH'').

To identify the source of the phase difference in the high-frequency
region $f \agt 400$\,Hz, we also plot a dot-dot line of $10(f/1\,{\rm
  kHz})^3 \propto x^{9/2}$ in Fig.~\ref{figap}. The slope of this
curve approximately captures the behavior of $\Phi_{\rm TT4}-\Phi_{\rm
  EOB}$ for $f \alt 1$\,kHz. Note that the phase in the TT4
approximant is calculated by
\beqn
\Phi_{\rm TT4}&=&2 \int x^{3/2}{dx \over (dx/dt)} \nonumber \\
&=& {5 \over 8} \int x^{-7/2} {dx \over F(x)},
\label{eq:phi}
\eeqn
where the right-hand side of Eq.~(\ref{eq:TT4}) should be substituted
for $dx/dt$ and $F(x)$ denotes the terms in $[\cdots]$ of
Eq.~(\ref{eq:TT4}): $1-487x/168 \cdots$. Equation~(\ref{eq:phi})
indicates that the error of $\Phi_{\rm TT4}$, which is associated with
the insufficiency for incorporating higher-order PN tidal effects,
should be of order $x^{4}$ for the 1.5PN tidal effect and $x^{9/2}$
for the 2PN tidal effect.  The slope of Fig.~\ref{figap} indicates
that the lack of such higher PN tidal effects would be the dominant
source of the disagreement.

Figure~\ref{figap} shows that the phase difference at $f \approx
400$\,Hz is appreciable; it is $\sim 0.2$, 0.3, and 0.4\,rad for the
DD2, TMA, and TM1 equations of state, respectively. This difference
results in disagreement of the hybrid-EOB and hybrid-TT4 waveforms as
illustrated in Sec. III. If the hybridization could be done for a
lower-frequency band, the phase difference would be smaller than $\sim
0.1$\,rad and the two hybrid waveforms would agree with each other in
a better manner.  However, Fig.~\ref{figap} suggests that the lack of
the more than 4PN non-tidal terms in the TT4 approximant would also
cause the phase disagreement of $O(0.1)$\,rad even for $f\alt 300$\,Hz
(if the coefficients of the 4PN terms were of $O(100)$, this would be
the case).  This lack could give non-negligible damage for making a
measurement template.  Higher-order non-tidal terms will be also
required for improving the TT4 approximant.

\section{Measurability in the Taylor-F2 approximant}

\begin{table*}
 \caption{$|| h_1 - h_2 ||$ in a TF2 approximant for a $1.35
   M_\odot$--$1.35 M_\odot$ binary at a hypothetical effective
   distance of $D_{\rm eff}=200$\,Mpc with several values of $f_i$ and
   $f_f$, which are shown in the upper-left corner of each table.
   ``$\Lambda=0$'' implies that the dimensionless tidal deformability
   $\Lambda$ employed is 0. * denotes the relation of symmetry.
\label{TF2}}
 \begin{tabular}{c|cccccc}
 ~10--500\,Hz\,\,  & $\Lambda=0$ & APR4 & SFHo & DD2 & TMA & TM1 \\
  \hline
  $\Lambda=0$  &---&0.6& 0.8& 1.6& 2.3& 2.7 \\
  APR4& * &---& 0.2& 1.0& 1.7& 2.1 \\
  SFHo& * & * & ---& 0.8& 1.5& 1.9 \\
  DD2 & * & * &  * & ---& 0.7& 1.1 \\
  TMA & * & * &  * &  * & ---& 0.5 \\
  TM1 & * & * &  * &  * &  * & ---
 \end{tabular}
\hspace{0.8cm}
 \begin{tabular}{c|cccccc}
  ~10--1000\,Hz \,\,& $\Lambda=0$ & APR4 & SFHo & DD2 & TMA & TM1 \\
  \hline
  $\Lambda=0$  &---& 1.5& 1.9& 3.2& 3.7& 4.1 \\
  APR4& * & ---& 0.5& 2.3& 3.2& 3.6 \\
  SFHo& * & *  & ---& 1.9& 3.0& 3.4 \\
  DD2 & * & *  & *  & ---& 1.5& 2.4 \\
  TMA & * & *  & *  & *  & ---& 1.1 \\
  TM1 & * & *  & *  & *  & *  & ---
 \end{tabular}
\\\vspace{5mm}
\begin{tabular}{c|cccccc}
10--2000\,Hz  & $\Lambda=0$ & APR4 & SFHo & DD2 & TMA & TM1 \\
  \hline
  $\Lambda=0$ & --- & 1.9 & 2.2 & 3.3 & 3.8 & 4.2 \\
  APR4 & * & --- & 0.7 & 2.5 & 3.3 & 3.7 \\
  SFHo & * & * & --- & 2.2 & 3.0 & 3.5 \\
  DD2 & * & * & * & --- & 1.8 & 2.5 \\
  TMA & * & * & * & * & --- & 1.3 \\
  TM1 & * & * & * & * & * & ---
 \end{tabular}
\hspace{0.81cm}
 \begin{tabular}{c|cccccc}
 30--2000\,Hz \,\,& $\Lambda=0$ & APR4 & SFHo & DD2 & TMA & TM1 \\
  \hline
  $\Lambda=0$  & --- & 1.8 & 2.1 & 3.1 & 3.7 & 4.0 \\
  APR4 & * & --- & 0.7 & 2.4 & 3.1 & 3.5 \\
  SFHo & * & * & --- & 2.1 & 2.9 & 3.3 \\
  DD2 & * & * & * & --- & 1.7 & 2.4 \\
  TMA & * & * & * & * & --- & 1.3 \\
  TM1 & * & * & * & * & * & ---
 \end{tabular}
\\\vspace{5mm}
 \begin{tabular}{c|cccccc}
50--2000\,Hz &$\Lambda=0$ & APR4 & SFHo & DD2 & TMA & TM1 \\
  \hline
  $\Lambda=0$ & --- & 1.7 & 2.0 & 2.9 & 3.4 & 3.7 \\
  APR4 & * & --- & 0.6 & 2.2 & 2.9 & 3.2 \\
  SFHo & * & * & --- & 1.9 & 2.7 & 3.1 \\
  DD2 & * & * & * & --- & 1.6 & 2.2 \\
  TMA & * & * & * & * & --- & 1.2 \\
  TM1 & * & * & * & * & * & ---
 \end{tabular}
\hspace{0.805cm}
\begin{tabular}{c|cccccc}
100--2000\,Hz & $\Lambda=0$ & APR4 & SFHo & DD2 & TMA & TM1 \\
  \hline
  $\Lambda=0$ & --- & 1.4 & 1.6 & 2.4 & 2.7 & 3.0 \\
  APR4 & * & --- & 0.5 & 1.8 & 2.3 & 2.6 \\
  SFHo & * & * & --- & 1.6 & 2.2 & 2.5 \\
  DD2 & * & * & * & --- & 1.3 & 1.8 \\
  TMA & * & * & * & * & --- & 0.9 \\
  TM1 & * & * & * & * & * & ---
 \end{tabular}
\end{table*}

By calculating $||h_1-h_2||$ of Eq.~(\ref{dSNR}), we also analyzed the
measurability of the dimensionless tidal deformability using a TF2
approximant of the inspiraling compact binaries of mass
$1.35M_\odot$--$1.35M_\odot$. Again, we employ the one-sided noise
spectrum density for the ``Zero Detuning High Power'' configuration of
advanced LIGO as $S_n (f)$~\cite{ligonoise}.  Here, for the TF2
approximant, we employ the spinless 3.5PN phasing~\cite{Ajith07}
incorporating the contribution of the tidal deformability up to 1PN
order with respect to the leading-order tidal
term~\cite{VFT11,wade14}. For the Fourier amplitude, we employ the 3PN
formulation for the point-particle approximation, described in
Ref.~\cite{khan15}, incorporating a tidal correction up to the 1PN
order~\cite{VFT11,damour12}.
Specifically, the spectrum is derived from a stationary phase 
approximation and is assumed to be written in a polynomial form:
\beqn
\tilde{h}_{\rm TF2}(f)&=&{m_0^2 \over D_{\rm eff}}
\sqrt{{5\pi \over 96}}\, (\pi m_0 f)^{-7/6} e^{i\psi_{\rm TF2T}(f)}
A_{\rm TF2T}(f), \nonumber \\
\eeqn
where
\beqn
A_{\rm TF2T}(f)&=&\sum_{k=0}^{12} A_k (\pi m_0 f)^{k/3}, \\
\psi_{\rm TF2T}(f)&=&2\pi f t_0 -\phi_0-{\pi \over 4} \nonumber \\
&+& {3 \over 32} (\pi m_0 f)^{-5/3} \sum_{k=0}^{12} \alpha_k (\pi m_0 f)^{k/3},
\label{eqTF2}
\eeqn
and the non-zero components of $A_k$ and $\alpha_k$ in our analysis are 
\beqn
&&A_0=1,~~A_2=-{37 \over 48},~~
A_4=-{9237931 \over 2032128},\nonumber \\
&&A_6={41294289857 \over 7510745088}-{205\pi^2 \over 192}
\nonumber \\
&&A_{10}=-{27 \over 16}\Lambda ,~~A_{12}=-{449 \over 64}\Lambda, \label{eqAA}
\eeqn
\beqn
&&\alpha_0=1,~~\alpha_2={2435 \over 378},~~\alpha_3=-16\pi, 
\nonumber \\
&&\alpha_4={11 747 195 \over 254016},~~
\alpha_5={9320 \over 189}\pi
\left[1+\ln(\pi m_0 f)\right], 
\nonumber \\
&&\alpha_6={1 382 467 552 339 \over 1 173 553 920} 
- {6848 \over 21}\gamma_{\rm E} - {7985 \pi^2 \over 48} \nonumber \\
&&~~~~~~~-{6848\gamma_{\rm E} \over 63}\ln(64\pi m_0 f) \nonumber \\
&&\alpha_7={1 428 740 \over 3969}\pi, \nonumber \\
&&\alpha_{10}=-{39 \over 2}\Lambda,~~\alpha_{12}=-{3115 \over 64}\Lambda. 
\eeqn
Here, $t_0$ is the coalescence time, $\phi_0$ is the coalescence
phase, $\gamma_{\rm E}$ is the Euler's constant, and $m_0$ is the
total mass.  We restrict our attention only to the formulation in the
equal-mass case.


The analysis for the measurability of the tidal deformability was
performed varying $f_i$ and $f_f$. Six results with different values
of $f_i$ and $f_f$ are listed in Table~\ref{TF2}. Here, we should pay
attention only to the results with $f_f$ smaller than $\sim 2$\,kHz
because by the TF2 approximant, the merger and post-merger waveforms
are not taken into account.  Comparing the results of $(f_i,
f_f)=(10\,{\rm Hz}, 2000\,{\rm Hz})$, $(30\,{\rm Hz}, 2000\,{\rm
  Hz})$, $(50\,{\rm Hz}, 2000\,{\rm Hz})$, and $(100\,{\rm Hz},
2000\,{\rm Hz})$, we find that employing $f_i=30$\,Hz, 50\,Hz, and
100\,Hz, the values of $||h_1-h_2||$ are systematically underestimated
by $\sim 5$\%, 15\%, and 30\%, respectively.

Comparing the results of $(f_i, f_f)=(10\,{\rm Hz}, 500\,{\rm Hz})$,
$(10\,{\rm Hz}, 1000\,{\rm Hz})$, and $(10\,{\rm Hz}, 2000\,{\rm
  Hz})$, it is found that the values of $||h_1-h_2||$ are
underestimated by a factor of 2 for $f_f=500$\,Hz. This is reasonable
because the tidal-deformation effect in phasing is accumulated in the
final inspiral orbits most significantly.  The values of $||h_1-h_2||$
for $(10\,{\rm Hz}, 1000\,{\rm Hz})$ are only by $\leq 0.2$ smaller
than those for $(10\,{\rm Hz}, 2000\,{\rm Hz})$ for the case that
$||h_1-h_2|| \geq 2.5$. On the other hand, for $||h_1-h_2|| \alt 2$,
the difference between the two cases can be 0.3--0.4. For such case,
it would be necessary to choose $f_f > 1$\,kHz.

\begin{figure*}[t]
\begin{center}
\includegraphics[width=84mm]{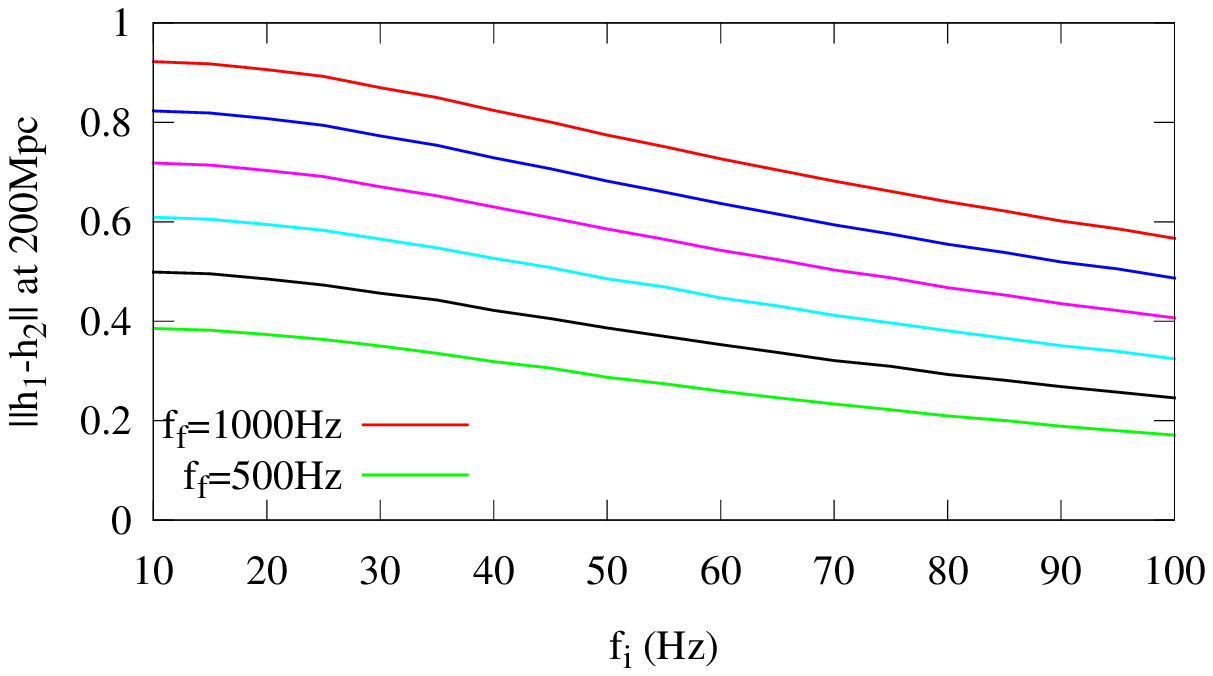}~~~
\includegraphics[width=84mm]{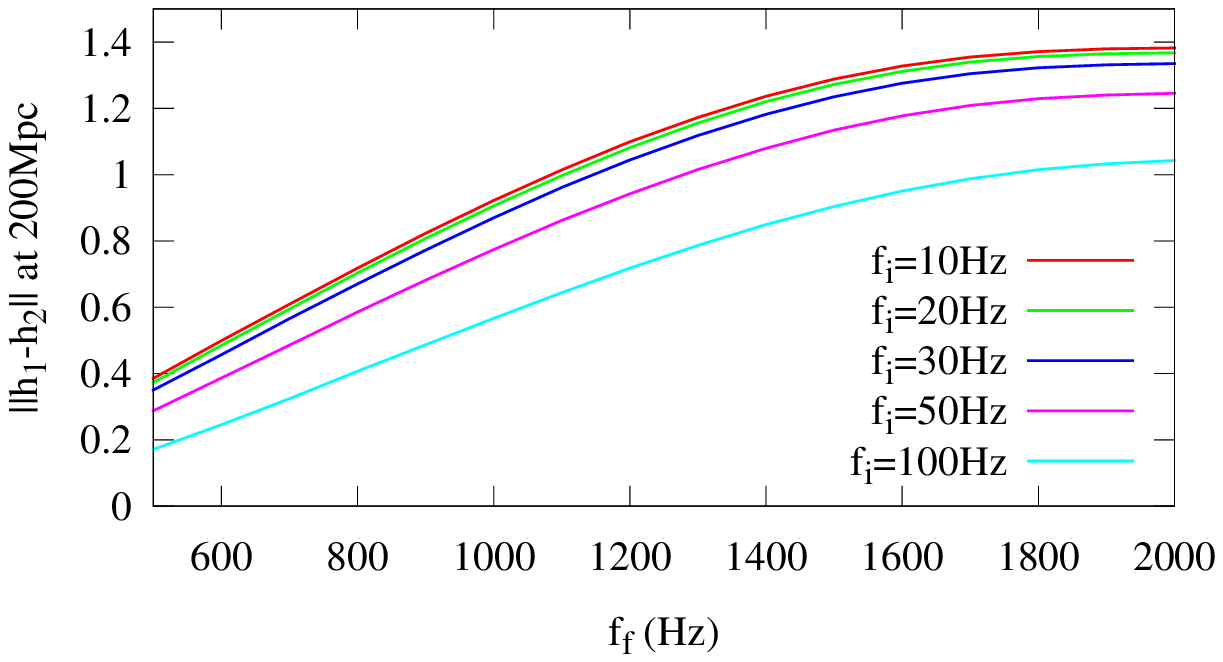}\\
\includegraphics[width=84mm]{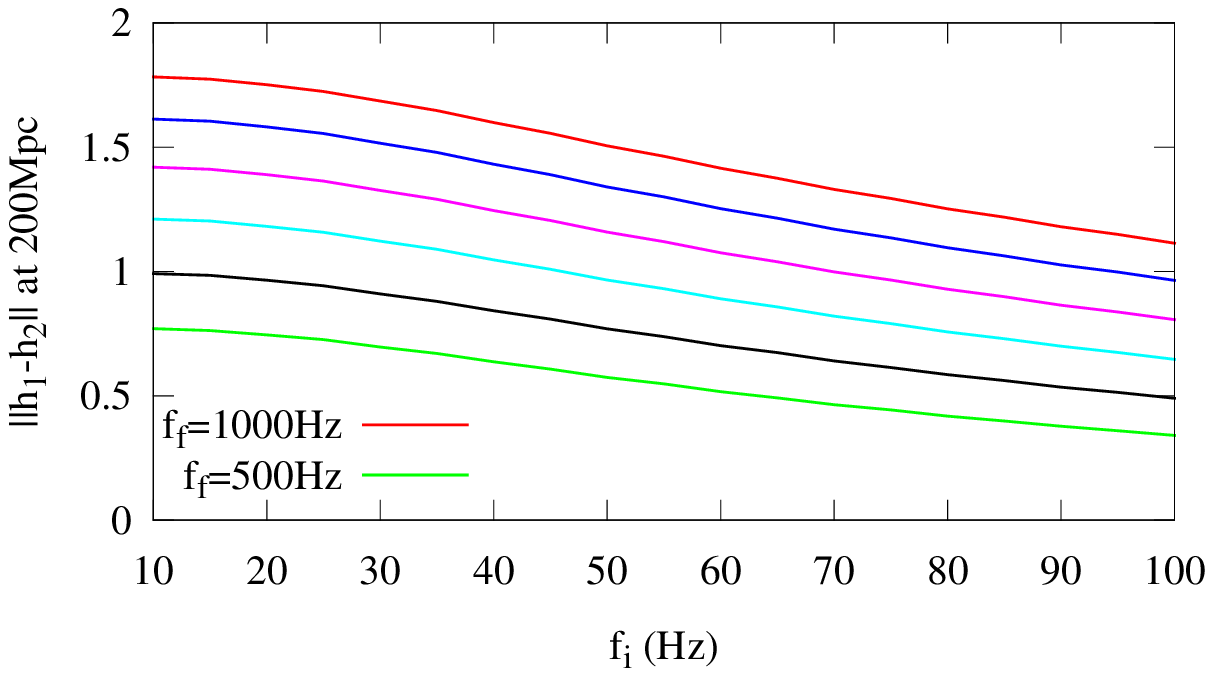}~~~
\includegraphics[width=84mm]{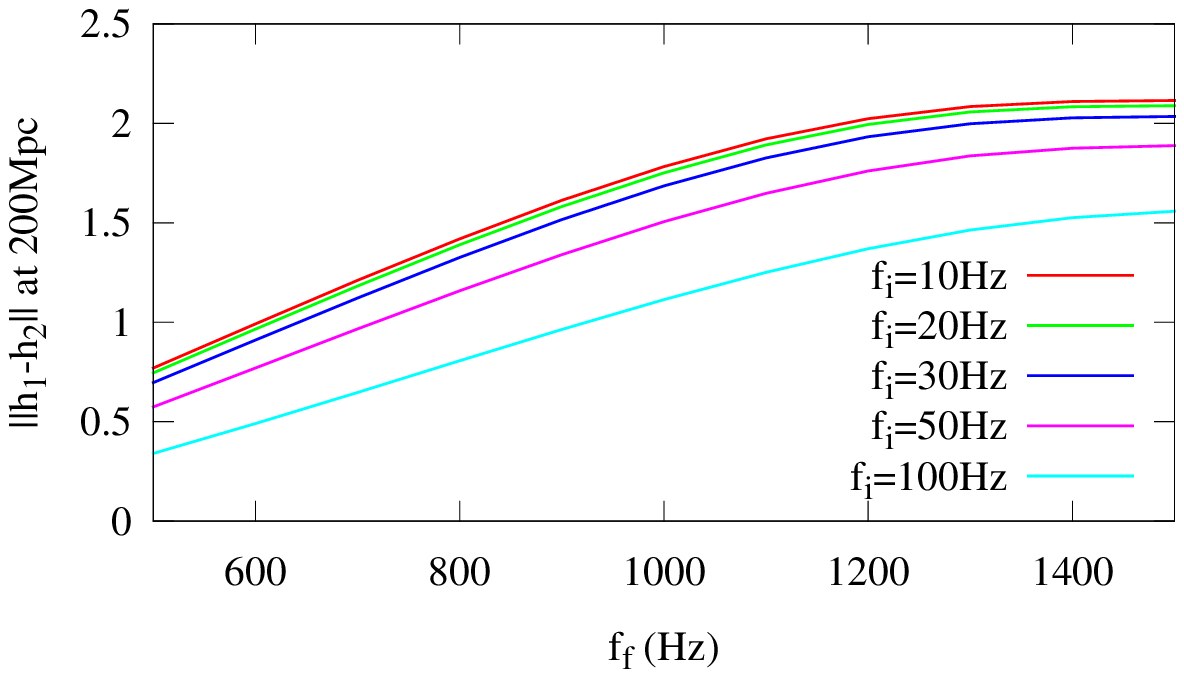}\\
\includegraphics[width=84mm]{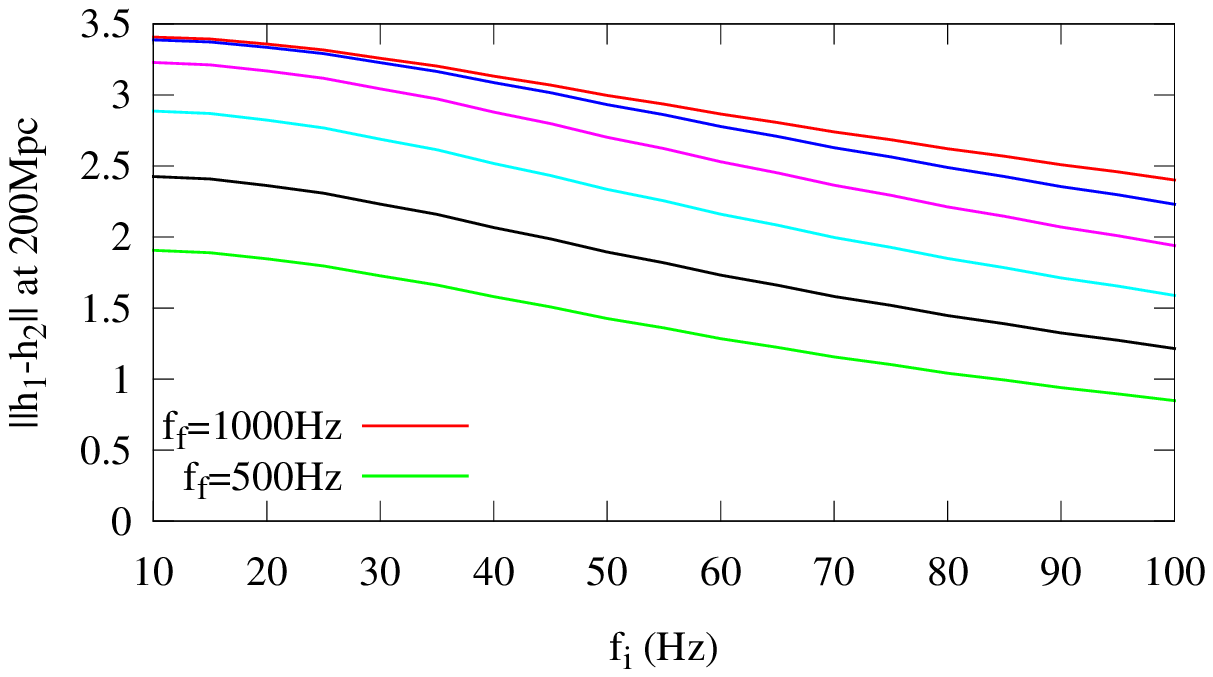}~~~
\includegraphics[width=84mm]{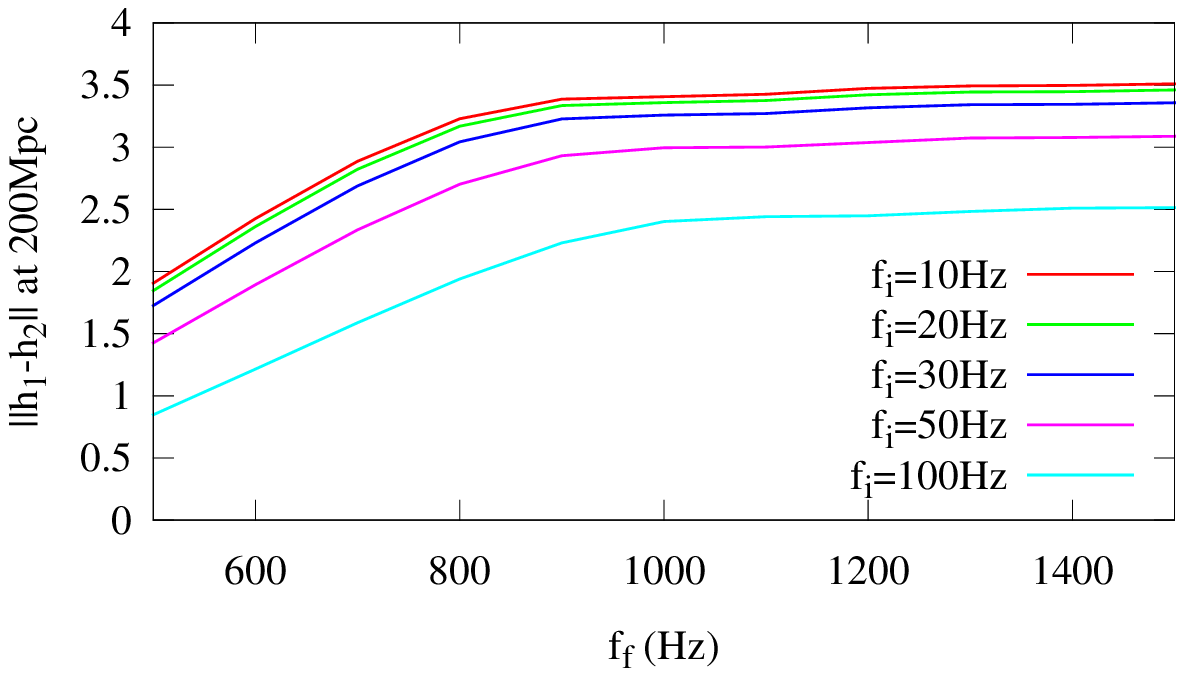}
\caption{Left panels: $||h_1-h_2||$ as a function of $f_i$ for
  $f_f=500$-- 1000\,Hz with $\delta \Lambda=200$ (top), 400 (middle),
  and 1000 (bottom), respectively. In each panel, the results for
  $f_f=500$, 600, 700, 800, 900, and 1000\,Hz are plotted (from the
  bottom to top curves). Right panels: $||h_1-h_2||$ as a function of $f_f$ for
  $f_i=10$-- 100\,Hz with $\delta \Lambda=200$ (top), 400 (middle),
  and 1000 (bottom), respectively.
\label{figap2}}
\end{center}
\end{figure*}

To further describe the dependence of $||h_1-h_2||$ on the choice of
$f_i$ and $f_f$, we generate Fig.~\ref{figap2}.  In the left three
panels of Fig.~\ref{figap2}, we plot $||h_1-h_2||$ at $D_{\rm
  eff}=200$\,Mpc as a function of $f_i$ for $f_f=500$--1000\,Hz with
$(\Lambda_1, \Lambda_2)=(200, 0)$ (top), $(400, 0)$ (middle), and
$(1000, 0)$ (bottom) (referred to as $\delta \Lambda=200$, 400, and
1000), respectively.  In the right three panels of Fig.~\ref{figap2},
we plot $||h_1-h_2||$ at $D_{\rm eff}=200$\,Mpc as a function of $f_f$
for $f_i=10$--100\,Hz with $\delta \Lambda=200$ (top), 400 (middle),
and 1000 (bottom), respectively.  As the left panels of this figure
indicate, the values of $||h_1-h_2||$ for $f_i=30$\,Hz and 50\,Hz are
respectively by $\approx 5$\% and $15\%$ smaller than those for
$f_i=10$\,Hz irrespective of $\delta \Lambda$ for which we choose a
realistic range.  This suggests that for $||h_1-h_2|| \alt 4$, the
values of $||h_1-h_2||$ are underestimated only for a small fraction
within 0.2 and 0.6, respectively, if we choose $f_i=30$\,Hz and
50\,Hz: Such fraction (in particular for $f_i=30$\,Hz) does not change
our conclusion in this paper. As the right panels indicate, this
property is independent of the choice of $f_f$.

The right three panels of Fig.~\ref{figap2} show that for a large
value of $\delta \Lambda \agt 400$, $||h_1-h_2||$ depends only weakly
on the choice of $f_f$ as long as it is larger than $\approx
1.5$\,kHz. For $\delta \Lambda = 200$, $||h_1-h_2||$ appears to
increase with $f_f$ even at $f_f=1.5$\,kHz. The reason for this is
that for a small value of $\delta \Lambda$, the values of
$||h_1-h_2||$ is accumulated relatively in a higher frequency range.
Thus for such case, it is necessary to take a high value of $f_f \sim
2$\,kHz: approximately the highest frequency of gravitational waves
prior to the merger.




\begin{thebibliography}{99}

\bibitem{Detectors} J.~Abadie {\it et al.}  
  Collaboration],
  Nucl.\ Instrum.\ Meth.\ A {\bf 624}, 223 (2010):
  T.~Accadia {\it et al.}  
  Class.\ Quant.\ Grav.\  {\bf 28}, 025005 (2011)
  [Erratum-ibid.\  {\bf 28}, 079501 (2011)]:
  K.~Kuroda, 
  Class.\ Quant.\ Grav.\  {\bf 27}, 084004 (2010).


\bibitem{aligo}Adcanced LIGO, http://www.advancedligo.mit.edu/

\bibitem{avirgo} Advanced VIRGO,\\
http://www.cascina.virgo.infn.it/advirgo/

\bibitem{kagra} KAGRA, http://gwcenter.icrr.u-tokyo.ac.jp/en/


\bibitem{0914} B. P. Abbott {\it et al.}, Phys. Rev. Lett. {\bf 116}, 
061102 (2016). 

\bibitem{Kalogera} V. Kalogera {\it et al.}, Phys. Rep. {\bf 442}, 75 (2007).

\bibitem{RateLIGO} J. Abadie et al. 
(The LIGO Scientific Collaboration and Virgo Collaboration), 
Class. Quantum Grav. {\bf 27}, 173001 (2010).

\bibitem{Kim} C. Kim, B. B. P. Perera, 
and M. A. McLaughlin, Mon. Not. R. Astro. Soc. {\bf 448}, 928 (2015). 

\bibitem{lattimer} J. M. Lattimer, Ann. Rev. Nucl. Part. Sci. {\bf 62}, 485 
(2012). 

\bibitem{lai94} D. Lai, F. A. Rasio, and S. L. Shapiro,
  Astrophys. J. \textbf{420}, 811 (1994).

\bibitem{mora04}T. Mora, and C. M. Will, Phys. Rev. D \textbf{69},
  104021 (2004).

\bibitem{flanagan08} E. E. Flanagan, and T. Hinderer, Phys. Rev. D
  \textbf{77}, 021502(R) (2008).

\bibitem{tania10}T. Hinderer, B. D. Lackey, R. N. Lang, and
  J. S. Read, Phys. Rev. D \textbf{81}, 123016 (2010).

\bibitem{VFT11} J. Vines, E. E. Flanagan, and T. Hinderer, Phys. Rev. D
  \textbf{83}, 084051 (2011).

\bibitem{BDF12} D. Bini, T. Damour, and G. Faye, Phys. Rev. D {\bf 85},
124034 (2012).

\bibitem{read13} J. S. Read, L. Baiotti, J. D. E. Creighton,
  J. L. Friedman, B. Giacomazzo, K. Kyutoku, C. Markakis, L. Rezzolla,
  M. Shibata, and K. Taniguchi, Phys. Rev. D \textbf{88}, 044042
  (2013).

\bibitem{pozzo13} W. Del Pozzo, T. G. F. Li, M. Agathos, C. Van Den Broeck, 
and S. Vitale, Phys. Rev. Lett. {\bf 111}, 071101 (2013).

\bibitem{favata14}M. Favata, Phys. Rev. Lett. \textbf{112}, 101101 (2014).

\bibitem{yagi14}K. Yagi and N. Yunes, Phys. Rev. D \textbf{89}, 021303 (2014).

\bibitem{wade14}L. Wade, J. D. E. Creighton, E. Ochsner, B. D. Lackey,
  B. F. Farr, T. B. Littenberg, and V. Raymond, Phys, Rev. D
  \textbf{89}, 103012 (2014).

\bibitem{agathos15} M. Agathos, J. Meidam, W. Del Pozzo, T. G. F. Li,
  M. Tompitak, J. Veitch, S. Vitale, and C. Van Den Broeck, Phys. Rev. D 
{\bf 92}, 023012 (2015).


\bibitem{thierfelder11} M. Thierfelder, S. Bernuzzi, and
  B. Br$\ddot{\rm{u}}$gmann, Phys. Rev. D \textbf{84},044012 (2011).

\bibitem{bernuzzi11} S. Bernuzzi, M. Thierfelder, and
  B. Br{\"u}gmann, Phys. Rev. D \textbf{85}, 104030 (2012).

\bibitem{bernuzzi12}S. Bernuzzi, A. Nagar, M. Thierfelder, and
  B. Br{\"u}gmann, Phys. Rev. D \textbf{86}, 044030 (2012).

\bibitem{HKS2013} K. Hotokezaka, K. Kyutoku, and M. Shibata, 
Phys. Rev. D {\bf 87}, 044001 (2013). 

\bibitem{radice14} D. Radice, L. Rezzolla, and F. Galeazzi,
Mon.\  Not.\ R.\ Astron.\ Soc.\ {\bf 437}, L46 (2014).

\bibitem{hotoke2015} K. Hotokezaka, K. Kyutoku, H. Okawa, and M. Shibata, 
Phys. Rev. D {\bf 91}, 064060 (2015). 

\bibitem{bernuzzi15} S. Bernuzzi, A. Nagar, T. Dietrich, and
  T. Damour, Phys. Rev. Lett. {\bf 114}, 161103 (2015). 

\bibitem{bernuzzi14} S. Bernuzzi, T. Dietrich, and A. Nagar, 
  Phys. Rev. Lett. {\bf 115}, 091101 (2015).

\bibitem{haas} K. Barkett, M. A. Scheel, R. Haas, C. D. Ott, 
S. Bernuzzi, D. A. Brown, B. Szil{\'a}gyi, J. D. Kaplan, J. Lippuner, 
C. D. Muhlberger, F. Foucart, and M. D. Duez, arXiv: 1509.05782. 

\bibitem{DN2010} T. Damour and A. Nagar, Phys. Rev. D {\bf 81}, 084016 
(2010).

\bibitem{damour12}T. Damour, A. Nagar, and L. Villain, Phys. Rev. D
  \textbf{85}, 123007 (2012).

\bibitem{DNB2013} T. Damour, A. Nagar, and S. Bernuzzi, Phys. Rev. D
  {\bf 87}, 084035 (2013). The code is available at
  https://eob-new.ihes.fr \,\,.

\bibitem{BD2013} D. Bini and T. Damour, Phys. Rev. D {\bf 87},
  121501(R) (2013).

\bibitem{BD2014} D. Bini and T. Damour, Phys. Rev. D {\bf 90}, 124037 (2014). 

\bibitem{KST2014} K. Kyutoku, M. Shibata, and K. Taniguchi, Phys. Rev. D 
{\bf 90}, 064006 (2014). 


\bibitem{yamamoto08} T. Yamamoto, M. Shibata, and K. Taniguchi, Phys. Rev. D
\textbf{78}, 064054 (2008).

\bibitem{BSSN} M. Shibata and T. Nakamura, Phys. Rev. D {\bf 52},
  5428(1995): T. W. Baumgarte and S. L. Shapiro, Phys. Rev. D {\bf
    59}, 024007(1998): M. Campanelli, C. O. Lousto, P. Marronetti, and
  Y. Zlochower, Phys. Rev. Lett. {\bf 96}, 111101 (2006): J. G. Baker,
  J. Centrella, D.-I. Choi, M. Koppitz, and J. van Meter,
  Phys. Rev. Lett. {\bf 96}, 111102 (2006).

\bibitem{Z4c} D. Hilditch, S. Bernuzzi, M. Thierfelder, Z. Cao, W. Tichy, 
and B. Br{\"u}gmann, Phys. Rev. D {\bf 88}, 084057 (2013). 

%

\bibitem{lorene} LORENE webpage: http://www.lorene.obspm.fr/  ~.

\bibitem{SFHo} A. Steiner, M. Hempel, and T. Fischer,
  Astrophys. J. {\bf 774}, 17 (2013).  

\bibitem{DD2} S. Banik, M. Hempel, and 
  D. Bandyophadyay, Astrophys. J.  Suppl. Ser. {\bf 214}, 22 (2014).

\bibitem{TM1} M. Hempel, T. Fischer, J. Schaffner-Bielich, and M.
  Liebend{\"o}rfer, Astrophys. J. {\bf 748,} 70 (2012).  

\bibitem{Shen} H. Shen, H. Toki,  K. Oyamatsu, and K. Sumiyoshi, 
Nucl. Phys. {\bf A637}, 435 (1998). 

\bibitem{demorest10} P. B. Demorest, T. Pennucci, S. M. Ransom,
  M. S. E. Roberts, and J.  W. T. Hessels, Nature \textbf{467}, 1081
  (2010): J. Antoniadis et al., Science, {\bf 340}, 448 (2013). 

\bibitem{APR4}A. Akmal, V. R. Pandharipande, and D. G. Ravenhall,
  Phys. Rev. C \textbf{58}, 1804 (1998).


\bibitem{Hotokezaka2013} K.~Hotokezaka, K.~Kiuchi, K.~Kyutoku,
  T.~Muranushi, Y.~-I.~Sekiguchi, M.~Shibata and
  K.~Taniguchi, Phys.\ Rev.\ D {\bf 88}, 044026 (2013).

\bibitem{RP2011} C. Reisswig and D. Pollney, Class. Quantum Grav. {\bf 28}, 
195015 (2011). 

\bibitem{LNZC2010} C. O. Lousto, H. Nakano, Y. Zlochower, and M. Campanelli, 
Phys. Rev. D {\bf 82}, 104057 (2010). 

\bibitem{Nakano15} H. Nakano, Class. Quantum Grav. {\bf 32}, 177002 (2015).

\bibitem{lackey14} B. D. Lackey, K. Kyutoku, M. Shibata, P. R. Brady,
  and J. L. Friedman, Phys. Rev. D {\bf 89}, 043009 (2014).

\bibitem{tania} However, see, T. Hinderer et al., arXiv:160200599. 

\bibitem{Boyle07} M. Boyle, D. A. Brown, L. E. Kidder, A. H. Mrou{\'e}, H.
P. Pfeiffer, M. A. Scheel, G. B. Cook, and S. A. Teukolsky, Phys. Rev.
D {\bf 76}, 124038 (2007). 

\bibitem{Ajith07} P. Ajith et al., arXiv: 0709.0093. 

\bibitem{Boyle08} M. Boyle, A. Buonanno, L. E. Kidder,
A. H. Mrou{\'e}, Y. Pan, H.  P. Pfeiffer, and M. A. Scheel, 
Phys. Rev.  D {\bf 78}, 104020 (2008).

\bibitem{ligonoise} https://dcc.ligo.org/cgi-bin/DocDB/ \\
ShowDocument?docid=2974

\bibitem{SXS} https://www.black-holes.org/waveforms/

\bibitem{abadie10}J. Abadie \textit{et. al.}, Class. Quantum
  Grav. \textbf{27}, 173001 (2010): C. Kim, B. B. P. Perena, 
and M. A. McLaughlin, Mon. Not. R. Astro. Soc. {\bf 448}, 928 (2015). 

\bibitem{clark14} J. Clark, A. Bauswein, L. Cadonati, H.-Th. Janka, C. Pankow, 
and N. Stergioulas, Phys. Rev. D \textbf{90},
  062004 (2014).

\bibitem{lee08} L. Lindblom, B. J. Owen, and D. A. Brown, 
Phys. Rev. D {\bf 78}, 124020 (2008). 

\bibitem{TS2010} K. Taniguchi and M. Shibata, Astrophys. J. Supplement 
{\bf 188}, 187 (2010). 

\bibitem{khan15} S Khan, S. Husa, M. Hannam, F. Ohme, M. P{\"u}rrer, 
X. J. Forteza, and A. Boh{\'e}, arXiv: 1508.07253. 


\end{thebibliography}
\end{document}